# Vortex-to-velocity reconstruction for wall-bounded turbulence via a data-driven model

Chengyue Wang [1, 3], Qi Gao [2, 3],†, Biao Wang [1],†, Chong Pan [3], Jinjun Wang [3]

[1] Sino-French Institute of Nuclear Engineering and Technology,

Sun Yat-sen University, Zhuhai, China

[2] School of Aeronautics and Astronautics, Zhejiang University, Hangzhou, China

[3] Key Laboratory of Fluid Mechanics, Ministry of Education,

Beihang University, Beijing, China

Modelling the vortex structures and then translating them into the corresponding velocity fields are two essential aspects for the vortex-based modelling works in wall-bounded turbulence. This work develops a data-driven method, which allows an effective reconstruction for the velocity field based on a given vortex field. The vortex field is defined as a vector field by combining the swirl strength and the real eigenvector of the velocity gradient tensor. The distinctive properties for the vortex field are investigated, with the relationship between the vortex magnitude and orientation revealed by the differential geometry. The vortex-to-velocity reconstruction method incorporates the vortex-vortex and vortex-velocity correlation information and derives the inducing model functions under the framework of the linear stochastic estimation. Fast Fourier transformation is employed to improve the computation efficiency in implementation. The reconstruction accuracy is accessed and compared with the widely-used Biot-Savart law. Results show that the method can effectively recover the turbulent motions in a large scale range, which is very promising for the turbulence modelling. The method is also employed to investigate the inducing effects of vortices at different heights, and some revealing results are discussed and linked to the hot research topics in wall-bounded turbulence.

________________________________________________________________________

## 1. Introduction

The concept of vortex is of prime importance in revealing the underlying physics behind complex flows. In wall-bounded turbulence, roles of vortices have been emphasized by extensive research works aiming to shed light on the self-sustaining mechanism of turbulence, to reveal the formation and evolution process of coherent structures and to develop a prediction theory for statistics. In various self-sustaining mechanisms as reviewed by Panton (2001), quasi-streamwise vortices are always closely linked to the low-speed streaks populated in the buffer layer (Schoppa & Hussain 2002), and they play a vital role in the autonomous cycle of near-wall turbulence (Jiménez & Pinelli 1999). The research works focusing on the outer layer give credit to the role of the hairpin vortices and their well-organized packets as the building blocks of large-scale motions (Adrian, Meinhart & Tomkins 2000). The hairpin model has also been chosen as the realization form for the attached eddy (Perry & Chong 1982; Perry & Marusic 1995), which was put forward to model the turbulence in logarithmic layer (Townsend 1957). The combination of hairpin model and attached eddy theory has achieved vast success in predicting the statistics (Woodcock & Marusic 2015; Yang, Marusic & Meneveau 2016a; Yang, Marusic & Meneveau 2016b) and in recovering the flow features (de Silva, Hutchins & Marusic 2016a; de Silva 2018) of the logarithmic region, which has been recently reviewed by Marusic & Monty (2019).

Inspired by these promising investigations, a number of researchers are devoted to the study on the characteristics of vortices. Tanahashi and his coworkers (Tanahashi *et al.* 2004; Das *et al.* 2006; Kang, Tanahashi & Miyauchi 2007) extensively investigated the properties of the tube-like vortices (termed as coherent fine-scale eddies, CFSE) based on the direct numerical simulations (DNSs) of turbulent channel flows. When normalized by local Kolmogorov scale ($\eta$) and the root mean square of velocity fluctuations ($u_{rms}$), the diameter and the maximum azimuthal velocity for CFSE are 10-12 $\eta$ and 0.5-0.6 $u_{rms}$ respectively, independent with the wall-normal positions and Reynolds numbers (Das *et al.* 2006). The universal scaling laws gained further supported by del Alamo *et al.* (2006) based on DNS of turbulent channel flow, and by Stanislas, Perret & Foucaut (2008) and Herpin *et al.* (2013) based on stereoscopic particle image velocimetry (SPIV) on the turbulent boundary flows for $Re_\tau$ up to 6868. Herpin *et al.* (2013) took into account for the orientation of vortices by conducting SPIVs in both streamwise-wall-normal plane and spanwise-wall-normal plane. They found that the PDFs of vortex radius and vorticity in the logarithmic layer can be fitted by a log-normal distribution in good qualities, and the fitting parameters show remarkably independent of Reynolds numbers. These investigations on CFSEs indicates some universal aspects of vortices, which provide some clues for the vortex-based modelling works.

Orientation is another aspect of vortex characteristics, which implies the elongating direction of vortex tubes and offers essential information to infer the topologies of dominant vortex structures. Ong & Wallace (1998), Ganapathisubramani, Longmire & Marusic (2006) and Kang *et al.* (2007) investigated the vortex orientation based on the inclination angles of vorticity vectors. One credit result is that the vortex structures at logarithmic layer tend to incline at an angle of 45° with respect to the streamwise direction, which had been theoretically explained by Head & Bandyopadhyay (1981). However, Bernard, Thomas & Handler (1993) and Gao, Ortiz-Duenas & Longmire (2007) showed that the local vorticity vector tends to be deflected away from the vortex axis at locations very close to the wall. Zhou (1997) demonstrated that the real eigenvector $\Lambda_r$ of the velocity gradient tensor is well aligned with the vortex axis by tracking $\Lambda_r$ within a hairpin eddy identified by swirl strength. Gao, Ortizdueñas & Longmire (2011) employed $\Lambda_r$ to characterize the vortex orientation and refined the vortex characteristics including orientations, circulation, propagation velocities at three different wall-normal positions based on both dual-plane PIV and DNS data. Recently, Tian *et al.* (2018) redefined the orientation of vortex by using a coordinate transformation and provided plenty of evidence that the newly-defined orientation is consistent with the vortex axis. Interestingly, Gao & Liu (2018) validated that the vortex orientation defined by Tian *et al.* (2018) is actually equivalent to $\Lambda_r$, which further promotes this identification criterion for vortex orientation. Using $\Lambda_r$ criterion, Wang *et al.* (2019) extended the work of Gao *et al.* (2011) and offered full information of vortex orientation, shapes and organizations for larger Reynolds number range ($Re_\tau = 1238 \sim 3081$) based on tomographic PIV data and DNS data.

Besides aforementioned works on general vortex tubes, plenty of attention has been put on some specially-selected vortices for the modelling purpose, such as the hairpin vortices, packets or attached structures. Christensen & Adrian (2001) statistically investigated the geometrical characteristics for packets structures in the streamwise-wall-normal plane, and found their streamwise inclination angle with respect to the wall is 12-13 degrees. Deng et al. (2018) filtered out the spanwise vortices not aligned in a packet via a POD-based filter technique and discovered a $\mathrm{Re}_\tau^{1/2}$ scaling of the saturated streamwise spacing between two hairpin heads in one packet. Wu & Christensen (2006) focused on the population trends of prograde and retrograde spanwise vortices in wall turbulence. They found that while the prograde spanwise vortices are populated in the inner boundary of the logarithmic layer, the retrograde ones are more prominent in the outer edge of the logarithmic layer. Natrajan, Wu & Christensen (2007) recognized the pairing trend of prograde and retrograde vortices by statistical methods and interpreted the vortex pairing as the signature of a $\Omega$-shape vortex. Besides these works limited in the streamwise-wall-normal plane, Ganapathisubramani, Longmire & Marusic (2003) recognized the patterns of vortex packets in the streamwise-spanwise plane and estimated that these patterns contribute more than 25% to the total Reynolds stress while occupy only 5% of the total area in the logarithmic layer. Hutchins, Hambleton & Marusic (2005) investigated the patterns of hairpin packets in the 135°-inclined plane with respect to the streamwise direction and found distinct two-regime behaviors: attached to the wall and detached from it, with a demarcation scaled well with the outer variables. The attached and detached structures were also reported by del Alamo et al. (2006), who recognized tall attached vortex clusters and detached vortex clusters based on 3D DNS data of turbulent channel flow. They found that the attached eddies are self-similar with a constant height-width-length ratio of 1:1.5:3 while the sizes of detached eddies are proportional to the local Kolmogorov scale.

These investigations have offered plenty of information about the size, intensity, orientation and organizations of vortices in wall-bounded turbulence, which are useful materials to validate and improve the existing vortex models. The pioneering works of Perry & Chong (1982) offered a promising implementation scheme for the attached eddy model, which took in the vortex geometry information and deduced statistics mathematically. This modelling scheme was further developed by Marusic (2001) who took into account for packet structures and estimated their roles for Reynolds stress and momentum transportation. Recently, the modelling scheme has been employed to recover the instantaneous flow features including the uniform momentum zone (de Silva et al. 2016a) and large coherence structures of spanwise velocities (de Silva 2018). These modelling works based on the vortex conception typically involves two necessary aspects: modelling the spatial organization of vortex and transfer the vortex distributions into velocity fields. While the former aspect could be built based on plenty of reports regarding universal scaling laws as listed above, less attention has been devoted to facilitate the later aspect. The explanation is that researchers tend to use the well-known Biot-Savart law for this purpose, which is prevailing for the vorticity-based reconstruction. For example, Perry & Marusic (1995) assumed Gaussian distribution of vorticity around the vortex axis and then employed the Biot-Savart law to recover the velocity fields. The argument for this reconstruction scheme is that their analysis is limited to the logarithmic region, where the deviation between vortex orientation and vorticity orientation is comparatively small (Wang et al. 2019). Nevertheless, things are different for the buffer layer because the orientation and distribution of vorticity and vortex are quite different therein. Extending application range for the vortex model to the buffer region will pose new challenges for the reconstruction technique, and Biot-Savart law may not be applicable. On the other hand, even if the Biot-Savart law works well in the logarithmic layer, it makes sense to seek for alternative reconstruction method and compare the accuracy with the Biot-Savart law.

A generalized discussion on how to reconstruct velocity fields based on given vortex fields add value to the well-developed vortex identification methods, which remains focused in the turbulence community up to present (Tian *et al.* 2018; Xiong & Yang 2019; Zhu & Xi 2019). The starting point of various vortex identification schemes is to extract tube-like vortices from complex flows, neglecting the shear layers embedded in the flow. The vortex tubes are quite sparse in the spatial field and could be simply represented as 3D curved lines with fewer adjustable parameters. Thus, the process of extracting vortices from turbulence is much like a data compression process as pointed out by Chakraborty, Balachandar & Adrian (2005), neglecting the incoherent or redundant information in turbulence. Vortex-to-velocity (V2V) reconstruction is the inverse process of this compression, which decompresses the vortex representations to recover original turbulence fields. Naturally, the compression and decompression are two fundamental aspects for the vortex-representing viewpoint of wall-bounded turbulence, yet only the former has been focused in the previous investigations.

V2V reconstruction offers a route to estimate the inducing role of vortices in turbulence quantitatively. A correlated work is reported by Pirozzoli, Bernardini & Grasso (2010), who compared the roles of vortex tubes and vorticity sheets (termed as vortex sheets in their article) by separately reconstructing velocity fields based on these two types of structures. A Poisson equation was solved to implement the reconstruction, which had been developed for the reconstruction regarding the intense vorticity (Jiménez *et al.* 1993). A divergence removing procedure was involved in order to make the field of vortex tubes or vorticity sheets more like a real vorticity field. The reconstructed results based on only vortex tubes are not very organized, especially at the buffer layer compared to the results of vortex sheets. They concluded that the vorticity sheets had a more important collective effect and contributed more to the turbulence production, which seems to violate the long-standing preconception that tube-like vortices take priority in the analysis of turbulent flows. However, it should be pointed out that the roles of vortex tubes might be underestimated since the reconstruction method is only strictly valid for the vorticity-based reconstruction. Developing a feasible method for V2V reconstruction would help to solve the controversy on the role of vortex tubes.

V2V reconstruction also benefits some vortex-based numerical simulation methods (Bernard 2009; Bernard, Collins & Potts 2010). The vortex methods track discrete vortex filaments in a Lagrangian frame, which provides an alternative to the grid-based simulation methods. In the vortex methods, the transformation from numerical vortex filaments to the velocity fields are accomplished by using a similar scheme as Perry & Marusic (1995) (as introduced before), and vorticity serves as an intermediate agent in the transformation. As a consequence, the resulting vortex filament must be aligned with the local vorticity line. Thus the numerical vortex filament is actually a vorticity filament, which is different from the vortex filament defined based on the swirl strength or other identification criteria. An effective V2V reconstruction scheme will allow a direct transformation from the criterion-identified vortex filaments to the velocity fields, which paves the way to develop a numerical simulation based on tracking real vortex filaments. One anticipated advantage for this simulation method is that it needs fewer discrete filament elements since the identified vortex field tends to be sparser than the vorticity field.

Motivated by these considerations, this work will focus on an effective V2V reconstruction technique. The V2V reconstruction will be implemented by a data-driven method termed as field-based linear stochastic estimation, which is a generalized method to estimate one 3D field based on another. Mathematically, the new method could be viewed as a least-square estimation on the linear inducing effect of vortices. The reconstruction accuracy will be estimated and compared with the Biot-Savart law. Before introducing the V2V reconstruction, this work will first discuss the properties of vortex as a vector field, with the purpose of shedding light on the distinct properties of the vortex field compared to the vorticity field. An interesting aspect about the relationship between vortex magnitude and orientation will be revealed based on the differential geometry, which adds value to the V2V reconstruction.

The following part of this work is arranged as follows. In Section 2, the DNS data employed in this investigation and associate numerical processing techniques are introduced. In Section 3, some novel aspects of the vortex field are introduced. In Section 4, the theory and implementation for field-based linear stochastic estimation as a V2V reconstruction method are introduced. In Section 5, the V2V reconstruction method is numerically validated and compared with the Biot-Savart law. The inducing role of vortices and the limitations of the method are further discussed based on the reconstruction results, followed by the concluding remarks in Section 6.

## 2. The DNS data and processing techniques

### 2.1 DNS data

The data employed in this work are acquired from an open-access DNS database (https://torroja.dmt.upm.es/) for high Reynolds number turbulent boundary layer (TBL). The numerical implementation for DNS has been reported by Borrell, Sillero & Jiménez (2013) and the more validations on the statistics are provided by Sillero, Jiménez & Moser (2013). A number of investigators (Marusic, Baars & Hutchins 2017; Wang, Wang & He 2017b; Wang *et al.* 2018) cited the DNS dataset in their investigation works on TBL. In our previous work (Wang *et al.* 2019), the conditionally-averaged vortex patterns extracted from this DNS data agree remarkably well with the results from tomographic PIV measurement, which inspires us to keep using the DNS data in the present work.

The whole DNS field corresponds to a developing zero pressure gradient TBL from $Re_\tau \approx 980$ to $Re_\tau \approx 2030$, with a total computation grid of $15361 \times 4096 \times 535$ for the streamwise, spanwise and wall-normal directions. In order to save memory, the separation between adjacent collocation points along the wall-normal direction is non-uniform, which is determined by local Kolmogorov scale (Borrell *et al.* 2013). The DNS dataset is fully-resolved and covers a very large dynamic range, which provides an ideal database for the implementation and validation of V2V reconstruction. In this work, only the DNS data segments for $Re_\tau = 1148 \sim 1250$ is employed and the averaged boundary thickness normalized by the wall unit (WU) $\delta^+ = 1200$. The prescription for this boundary thickness is based on two considerations. From the perspective of V2V reconstruction, the resulting logarithmic layer covers a wall-normal range of about 160 wall units (from 80 WU to $0.2\delta^+$), which corresponds to an acceptable number of computational nodes for V2V reconstruction. On the other hand, plenty of reach works (Tanahashi *et al.* 2004; del Alamo *et al.* 2006; Herpin *et al.* 2013; Wang *et al.* 2019) has validated that the fine-scale vortices which are closely linked to the current work present universal behaviours, which are independent with Reynolds number. The truncated DNS data segments have a streamwise length of $8\delta^+$, a spanwise length of $13.4\delta^+$, and a wall-normal range below $0.3\delta^+$, corresponding to a Cartesian grid of $1412\times 4096\times 91$. The variation of the viscous wall unit in the DNS data segments is less than 0.4%, which is small enough to be viewed as a constant. Totally, twelve frames of DNS fields are downloaded, providing plenty of data to ensure fully-converged statistical results. To avoid the resolution issue and minimize the processing errors, all the derivative computations in the present work are based on the original DNS grid via a 5-nodes central differential scheme. Parameters about the DNS data segments employed in this work are collected in table 1.

Throughout this article, the coordinating system is built so that the origin is located on the wall, and $x, y, z$ align with the streamwise, spanwise and wall-normal direction, respectively. **u** denotes the fluctuating velocity vector, which is obtained by subtracting the mean streamwise velocity component from the original DNS vector field. $u, v, w$ designate the streamwise, spanwise and wall-normal fluctuating velocity components, respectively. A superscript of '+' indicates the quantity is normalized by the wall unit or the friction velocity. In this work, only the fluctuating velocity field is considered in the following analysis.

| Reynolds number range | Wall unit variation | Streamwise range (WU) | Spanwise range (WU) | Wall-normal range (WU) |
|---|---|---|---|---|
| $\mathrm{Re}_\tau = 1148 \sim 1250$ | ±0.4% | $8\delta^+$ | $13.4\delta^+$ | $4.99 \sim 359.3$ |
| Total number of nodes | Total number of frames | Streamwise spacing (WU) | Spanwise spacing (WU) | Wall-normal spacings (WU) |
| $1412\times 4096\times 91$ | 12 | 6.80 | 3.93 | $1.01 - 6.26$ |

TABLE 1. A collection of parameters for the DNS data employed in this work

*2.2 Vortex identification*

Various vortex identification methods have been developed to extract vortical structures in turbulence. Most of them are based on local velocity gradient tensor ($\nabla \mathbf{u}$), including the second invariant $Q$ (Hunt, Wray & Moin 1988), the discriminant $\Delta$ (Chong, Perry & Cantwell 1990), the swirl strength $\lambda_{ci}$ (Zhou *et al.* 1999), the $\lambda_2$ criterion (Jeong & Hussain 1995) and so on. Chakraborty *et al.* (2005) revealed the relationships between these local identification criteria and validated that results from different identification criterion are approximately equivalent for turbulent flows.

In these works, the vortex field was recognized as a scalar field, and certain thresholds should be prescribed when showing the iso-surface of vortex structures. The vortex field was not regarded as a vector field until Gao *et al.* (2011) and Tian *et al.* (2018). They recognized the vortex orientation by the real eigenvector of $\nabla \mathbf{u}$ ($\mathbf{\Lambda}_r$). Tian *et al.* (2018) gave a mathematical discussion on the definition of the vortex vector. According to the viewpoints of Tian *et al.* (2018), the vortex orientation ($\mathbf{\Lambda}_r$) corresponds to a direction with zero rotation speed.

Based on the definition of velocity gradient tensor ($\nabla \mathbf{u}$), the relative velocity $\delta \mathbf{u}$ for two points separated by a small position vector $\delta \mathbf{r}$ could be expressed as

$$\delta \mathbf{u} = \nabla \mathbf{u} \cdot \delta \mathbf{r}. \tag{2.1}$$

The non-rotate requirement means that the relative velocity should be along $\delta \mathbf{r}$, which means

$$\nabla \mathbf{u} \cdot \delta \mathbf{r} = \mu \delta \mathbf{r}, \tag{2.2}$$

where $\mu$ should be a real constant. The above equation implies that $\delta \mathbf{r}$ is the real eigenvector of $\nabla \mathbf{u}$, i.e. $\boldsymbol{\Lambda}_r$.

Considering $\boldsymbol{\Lambda}_r$ contains no sign information about rotation, the local vorticity $\boldsymbol{\omega}$ is employed as a reference. Specifically, the direction of $\pm \boldsymbol{\Lambda}_r$ corresponding to an acuate included angel with $\boldsymbol{\omega}$ is viewed as the vortex orientation. In this work, for the convenience of numerical implementation, the vortex intensity $\lambda_{ci}$ and orientation vector $\boldsymbol{\Lambda}_r$ are combined to define a vector field as

$$\boldsymbol{\Lambda} = \lambda_{ci} \boldsymbol{\Lambda}_r \frac{\boldsymbol{\Lambda}_r \cdot \boldsymbol{\omega}}{|\boldsymbol{\Lambda}_r||\boldsymbol{\omega}|}. \tag{2.3}$$

Throughout this article, the vortex field refers to $\boldsymbol{\Lambda}$ as default, which is consistent with Wang *et al.* (2019).

### 2.3 Linear coherent spectrum

Linear coherent spectrum (LCS) is a diagnostic tool for the scale-by-scale coupling of two input signals (Bendat & Piersol 1986) denoted as $u_1(x), u_2(x)$. The formula for LCS takes a form like the correlation coefficient as

$$\text{LCS}(\kappa) = \frac{\langle \tilde{u}_1(k) \tilde{u}_2(k)^* \rangle}{\sqrt{\langle \tilde{u}_1(k) \tilde{u}_1(k)^* \rangle \langle \tilde{u}_2(k) \tilde{u}_2(k)^* \rangle}}, \tag{2.4}$$

where $\tilde{u}_1(k)$ and $\tilde{u}_2(k)$ denotes the Fourier transform coefficients corresponding to the wavenumber of $k$ and the superscript star designates the complex conjugate. $\langle \cdot \rangle$ represents an ensemble average which could be implemented by averaging along the homogenous dimensions in wall-bounded turbulence.

What is noteworthy is that the above equation is slightly different from that of Baars & Marusic (2020) by abandoning the modulus operation on the numerator, which results in a complex value of LCS. By definition, the intention of LCS is to decompose the input signals into sinusoidal functions by Fourier transform and then estimate the wavenumber-specific coherent degrees of these sinusoidal components. The *magnitude* of LCS indicates the correlation coefficient for two scale-specific sinusoidal components with phase consistency, while the *phase angle* of LCS indicates the relative phase angle or the corresponding phase deviation. In this work, both the magnitude and phase angle of LCS are employed to assess the reconstruction accuracy of velocity fields.

## 3. New aspects of the vortex field

This section will discuss the distinctive properties of the vortex field as defined by Section 2.2. The differential geometry will be employed to shed light on the relationship between vortex magnitude and orientation. Subsequently, the divergence property will be examined and evaluated by using a divergence correction scheme. These two properties will be combined in an attempt to infer the three vortex components based on the vortex magnitude, followed by a further discussion extended to the topic of the V2V reconstruction.

*3.1 The relationship between the vortex magnitude and orientation*

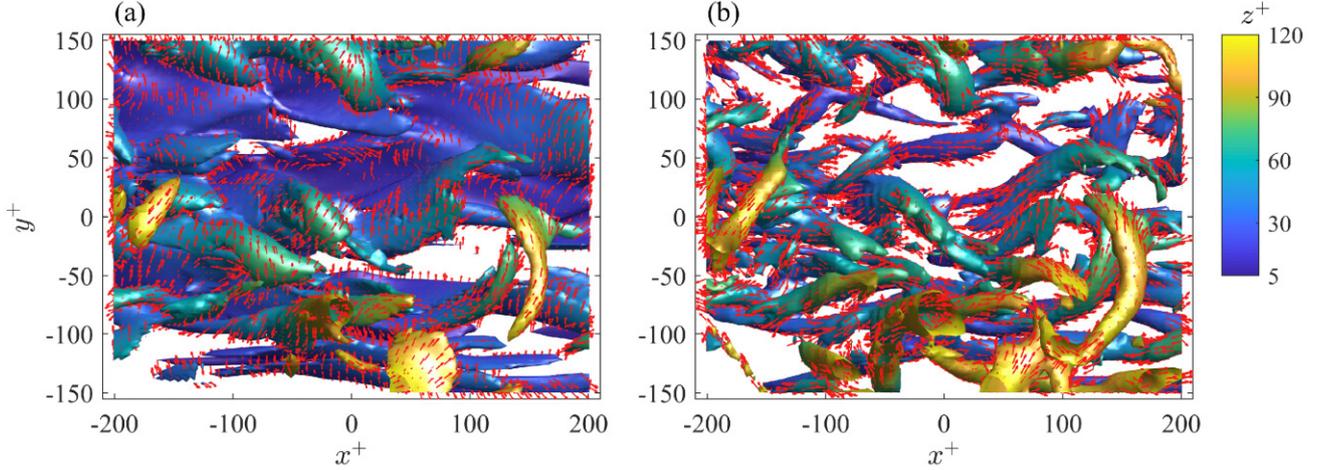

FIGURE 1. Instantaneous flow structures identified by vorticity magnitude (a) and vortex magnitude (b). The surfaces in the two plots correspond to $\omega^+ = 0.42$ and $\lambda_{ci}^+ = 0.10$ respectively, coloured based on the distance from the wall. Orientations of vorticity and vortex on the surfaces are also displayed by red vectors. For clarity, only one out of four vectors in each direction are shown.

To start the discussion about the vortex magnitude and orientation, an observation on the instantaneous field is beneficial. Figure 1 shows the structures identified by isosurfaces of vorticity magnitude ($\omega$) and vortex intensity ($\lambda_{ci}$), with orientation vectors displayed on the surfaces. Since both $\omega$ and $\lambda_{ci}$ are calculated based on the fluctuating velocity field, the influence of mean shear in TBL has been removed. The thresholds for $\omega$ and $\lambda_{ci}$ are prescribed so that the identified structures occupy about 10% of the total volume considered. Distinct characteristics for the $\omega$-identified and $\lambda_{ci}$-identified structures are noticed. The former contains large bulks of flat sheet-like entities at the lower layer and thin tube-like entities above. The later recognizes only the vortex tubes with a thickness of about ten wall units. From the perspective of modelling, tube-like structures are easier for mathematical representation than sheet-like structures since they could be simplified as 3D curves with varying radius. Another convenience for modelling vortex fields is that the vortex vectors are mostly attached in the vortex surface and pointing to the extending direction of the vortex tubes. This property has also been reported by Tian *et al.* (2018), who regarded it as a proof for their definition of vortex orientation. For the vorticity field, the property does not hold at the near-wall region, where the vorticities tend to be aligned with the spanwise direction. A correlated work involving this property is reported by Xiong & Yang (2019), who attempted to calculate the "vorticity surface field" based on the vorticity field. A scalar field is called the "vorticity surface field" if the normal vector of its iso-surface is perpendicular to the local vorticity vector. If this formulation is employed to define a "vortex surface field", $\lambda_{ci}$ could be regarded as the vortex surface field as we can see from figure 1.

To quantitatively analyze the relationship between the vortex surface and the vortex orientation, we resort to the differential geometry, which is a useful mathematical tool to describe the local curvatures of a surface. Consider a small vortex surface element $S$ as shown in figure 2(a), small enough to be viewed as a quadratic surface. $\mathbf{n}$ denotes a normal vector of $S$, pointing to the side with smaller vortex magnitude. Generally, two orthogonal principal directions (indicated by two unite vectors: $\mathbf{\kappa}_1$ and $\mathbf{\kappa}_2$) can be determined on $S$, which correspond to the directions with the maximum and minimum curvatures, respectively. $\mathbf{\kappa}_1$ and $\mathbf{\kappa}_2$ refer to the first principal direction and the second principal direction for the local curvatures. Taking an example of a straight vortex tube with a uniform radius (see figure 2(b)), the two principal directions ($\mathbf{\kappa}_1$ and $\mathbf{\kappa}_2$) for the local curvature are aligned with the axial direction and azimuthal direction, respectively. The corresponding curvatures along $\mathbf{\kappa}_1$ and $\mathbf{\kappa}_2$ refer to the local curvatures of lines extracted along the two directions from the surface (see $l_1$ and $l_2$ in figure 2(b)), which are 0 and $-1/R$ respectively. The negative sign indicates that the extracted line ($l_2$) bends toward the negative direction of $\mathbf{n}$. It can be inferred that for a general tube-like vortex surface, the curvature along $\mathbf{\kappa}_2$ is always negative. One can guess that the vortex orientation tends to align with $\mathbf{\kappa}_1$, which will be tested and verified in the flowing discussion.

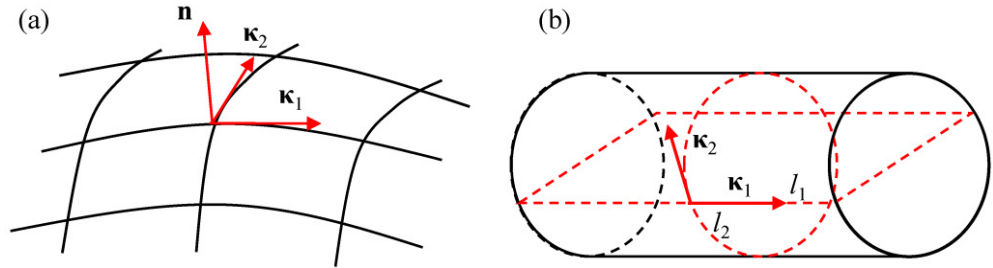

FIGURE 2. Diagrammatic sketches for the principal directions of the local curvatures

For each point in the vortex field, the vortex surface passing this point could be extracted, and correspondingly the wall-normal vector $\mathbf{n}$ and the principal directions ($\mathbf{\kappa}_1$ and $\mathbf{\kappa}_2$) can be determined. Specific description for this process is provided in Appendix A of this article. $\mathbf{\kappa}_1$, $\mathbf{\kappa}_2$ and $\mathbf{n}$ constitute a local coordinate system, and we will statistically investigate the vortex orientation in this local coordinate system. Specifically, a coordinate transformation from the global coordinate system to this local coordinate system is performed on the vortex vector, which results in a new vector form denoted as $\mathbf{\Lambda}'$. To investigate its orientation, $\mathbf{\Lambda}'$ is further expressed by three spherical coordinates: radical distance ($\rho$), azimuth angle ($\psi$) and zenith angle ($\theta$). The joint PDF of $\psi$ and $\theta$ for $\mathbf{\Lambda}'$ are calculated based on the data points for $\lambda_{ci} > 0.10$, which has been employed as the threshold in figure 1 and is deployed here to exclude the influence of weak vortices. The resulting PDF is further normalized by $\sin(\theta)$ before displayed on the sphere surface (figure 3) in order to reflect the local areal density of probability, which is called a spherical PDF. Unlike the joint PDF of $\psi$ and $\theta$ widely employed (Stanislas *et al.* 2008; Gao *et al.* 2011), the spherical PDF treat all the orientations equally and is covariant with the rotation of the coordinate system, which is beneficial for an objective judgement.

Taking the example of a completely random vector, the spherical PDF gives a uniform distribution on the sphere while the joint PDF of $\psi$ and $\theta$ shows a concentration at $\theta=90°$.

Returning to figure 3, only a quarter of a complete sphere is contoured since $\kappa_1$, $\kappa_2$ could change direction by 180°, blurring the definitions of $\Lambda'$ in these four quarters. To provide a reference, the spherical PDF for vorticity orientations in the curvature coordinates from the vorticity magnitude is also provided. The spherical PDF for vortex orientation shown in figure 3(b) presents an extremely dense distribution along the $\kappa_1$ direction. Specifically, the largest probability density appearing nearby $\kappa_1$ axis is about 6.8, which is 21 times as large as the average probability on the sphere quarter ($1/\pi$). It verifies that the vortex vector has a strong disposition to be aligned with the first principal direction of vortex surface curvature. The same relation does not hold for vorticity, which could align with both $\kappa_1$ and $\kappa_2$ with large probability, as shown in figure 3(a). What is noteworthy is that the first principal direction for curvature also correspond to the direction along which vortex magnitude change the least. Thus we can restate that the vortex orientation tends to be aligned with the slowest-changing direction of the vortex magnitude. The discussion in this section reveals an interesting aspect of the relationship between the vortex orientation and the vortex magnitude. This property indicates that the swirl strength $\lambda_{ci}$ and the eigenvector $\Lambda_r$ are kind of harmonious and thus combining them to define a vortex vector as we did in this work is quite reasonable.

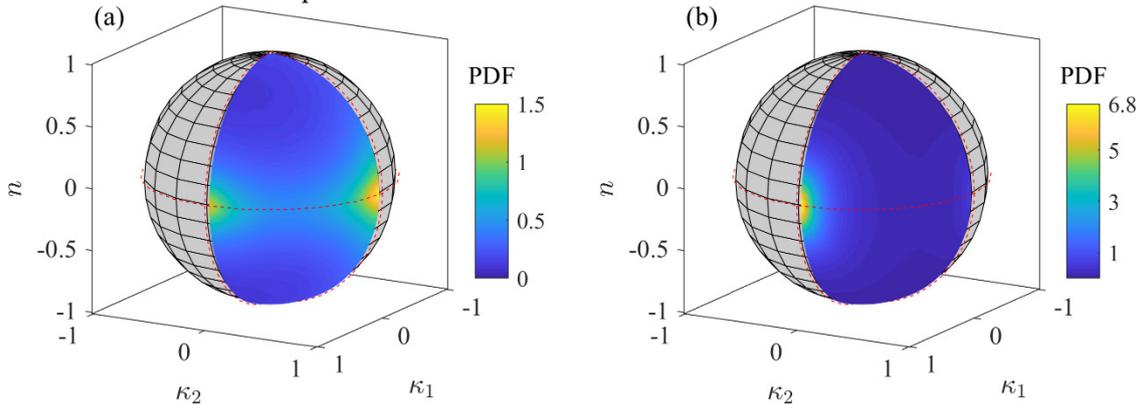

FIGURE 3. The spherical PDFs for $\boldsymbol{\omega}$ (a) and $\boldsymbol{\Lambda}$ (b) orientations in the coordinate system of the local curvatures. Note that the colour scales in the two subplots are different for better visualization.

*3.2 Divergence of the vortex field*

Vorticity field is naturally divergence-free based on its definition. The divergence property for vortex field will be examined in this work. A joint PDF for $\partial\Lambda_x/\partial x+\partial\Lambda_y/\partial y$ and $-\partial\Lambda_z/\partial z$ is displayed in figure 4, which is a popular way to check the divergence of a given vector field (Jodai & Elsinga 2016). The contours show a pattern of flat ellipses inclined along the diagonal direction, indicating that the vortex field tends to be divergence-free in a statistical sense. To quantify the distance from $\boldsymbol{\Lambda}$ to the nearest divergence-free field, the minimum correction scheme developed by Wang *et al.* (2017a) is employed to remove the divergence of $\boldsymbol{\Lambda}$ artificially. The deviation percentages (DPs) between original vortex fields and the corrected vortex fields $\boldsymbol{\Lambda}_c$ are defined as the 2-norm of the difference between the two fields, normalized by the 2-norm of $\boldsymbol{\Lambda}$. Consistent with former discussion, only the data for recognized vortices ($\lambda_{ci}>0.10$) and for a fixed wall-normal position from an instantaneous DNS field (introduced in table 1) are counted in the statistics. DPs are plotted as a function of wall-normal position, as shown in figure 4(b). It shows that throughout the wall-normal range considered, the average DP is less than 10%. At the region very close to the wall ($z^+ < 20$), the DP is comparatively large. For $z^+ > 20$, the largest deviation is about 12% occurring at $z^+ \approx 40$, above which the deviation continuously reduces. The volume fraction of recognized vortices as a function of $z$ is also displayed in the same plot. The largest volume fraction occurs at $z^+ \approx 30$, which implies the densest crowd of vortices. The vortex crowd seems to cause larger divergence errors, considering that the position of the DP peak is close to $z^+ \approx 30$. These results support the argument that the vortex field can be approximately regarded as a divergence-free field.

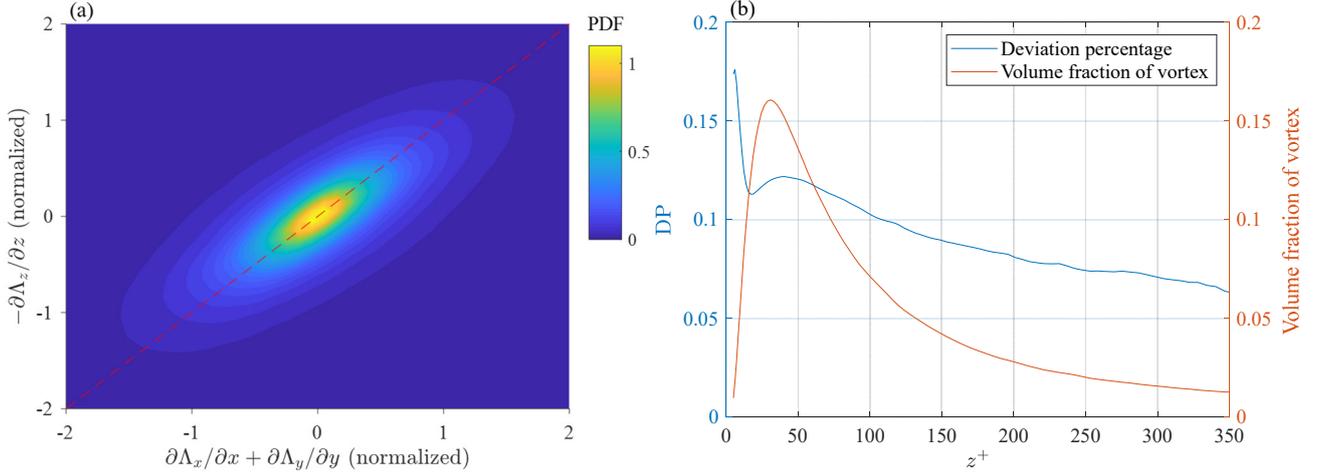

FIGURE 4. Results for the divergence testing on the vortex field. (a) Joint PDF for $\partial\Lambda_x/\partial x + \partial\Lambda_y/\partial y$ and $-\partial\Lambda_z/\partial z$, normalized by their standard deviations; (b) the deviation percentage between the original vortex field and the nearest divergence-free field as a function of $z^+$; the volume fraction for identified vortices are also plotted as a reference.

### 3.3 Inferring three vortex components based on the vortex magnitude

Employing the properties of $\Lambda$ introduced before, we can infer all the three vortex components based on the vortex magnitude. Specifically, the first curvature direction ($\kappa_1$) is calculated based on the differential geometry. The local vorticity is used as a reference to deal with the sign issue of $\kappa_1$, as we do in the definition of $\Lambda$. The magnitude $\lambda_{ci}$ and $\kappa_1$ are combined to define the vortex field, followed by the divergence-free correction scheme to improve the results. The inferred vortex components for $z^+ = 50.9$ are displayed in figure 5, compared with the ground truth DNS field. As we can see, the inferred results are remarkably consistent with the ground truth ones. Outliers are also noticed in the inferred results, which might be caused by the spurious local curvatures of vortex surfaces. An outlier correction scheme will further improve the current inferred result, which is not involved in this work. The correlation coefficients (CCs) for the original vortex fields and the inferred ones with or without divergence correction are plotted in figure 6 as a function of wall-normal positions. It shows the inferred vortex fields resemble the original vortex fields to the most extent for $z^+ \approx 20$. With the increase of $z$, CC first reduces and then level off at $z^+ > 100$. Divergence correction improves CCs by 0.05, which indicates a successful application for the divergence-free property of vortex fields.

Before this work, the vortex field is mostly viewed as a scalar field. This work expands the vortex field to be a vector field since it is necessary for V2V reconstruction. Generally, more information is needed to represent a vector field, which seems to violate the original intention of trying to compress redundant information for vortex identification. The analysis in this section helps to relieve this concern by validating that the orientation of vortex and its magnitude is not completely independent. In fact, it is possible to infer the orientation information based on the local magnitude information.

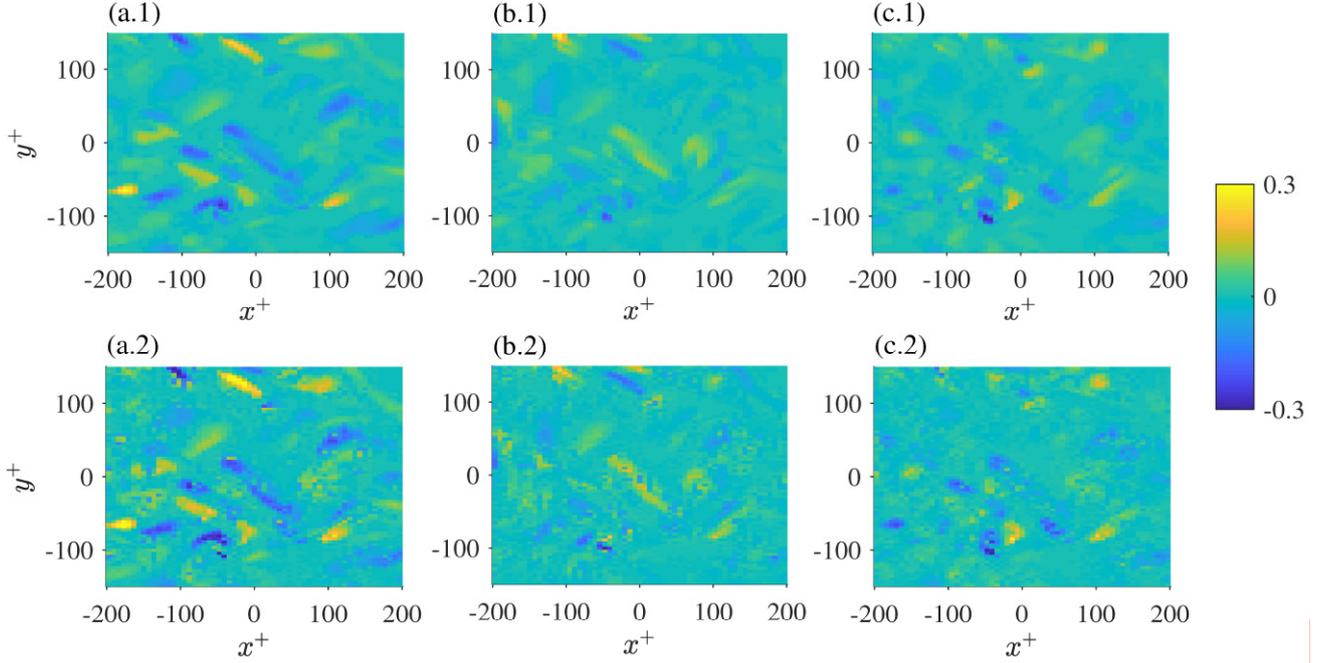

FIGURE 5. The original vortex field (the first row) and the inferred vortex field (the second row) at $z^+ = 50.9$. From left to right, the contours correspond to $\Lambda_x^+$, $\Lambda_y^+$ and $\Lambda_z^+$, respectively.

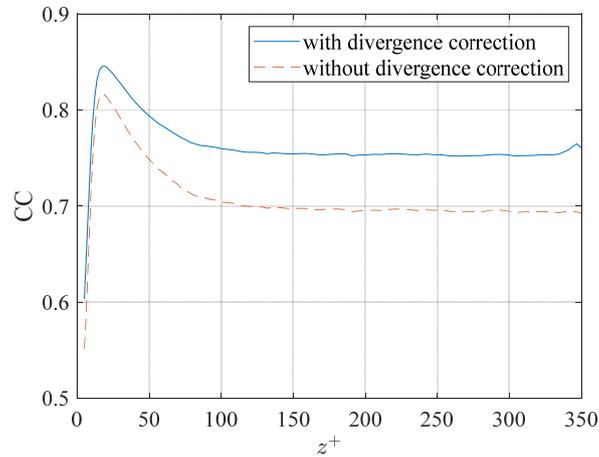

FIGURE 6. The correlation coefficients (CCs) for original vortex fields and inferred vortex fields with (solid line) or without (dash line) divergence correction as functions of $z^+$.

Results in this section facilitate a further discussion on the theoretical positions of vortex and vorticity in turbulence modelling. Vorticity has been regarded as an alternative for the presentation of vortex structures in turbulence (Bernard 2011, 2019; Xiong & Yang 2019). One advantage of vorticity is that it has a strict mathematical definition. For impressible flow, the velocity field and vorticity field are mathematically equal in the sense that if one is known, the other could be deduced precisely. The Biot-Savart law serves as an effective tool to transform the vorticity field into the corresponding velocity field. However, the velocity-vorticity equivalence indicates that the vorticity field is just as complex as the original velocity field, which eludes the vorticity-based modelling works. Different from the vorticity field, the vortex field (as defined by $\Lambda$ in this work) only focuses on the rotational motions, which could be simplified as the arrangement of tube-like geometries. Behaviours of these rotational motions are managed by more strict regulations, as revealed by the differential geometry in this work. The vortex fields could be determined by less information, providing more convenience for the corresponding modelling work. Furthermore, large numbers of research works (Tanahashi *et al.* 2004; del Alamo *et al.* 2006; Stanislas *et al.* 2008; Wang *et al.* 2019) have revealed the universal behaviours of vortices in the radius, intensity, orientation and correlation functions, further consolidating the position of vortices as ideal modelling objects. Nevertheless, everything has two sides. The vortex fields contain only the information of strong rotational motions, neglecting the vorticity sheets embedded in wall-bounded turbulence. Such information loss poses a challenge for the vortex-based reconstruction, which will be focused in the next section.

## 4. Vortex-to-velocity reconstruction

The primary challenge for V2V reconstruction is the information loss resulted from the vortex identification process. To pad the information gap, some empirical parameters or data-driven models should be employed in the V2V reconstruction. On the other hand, we hope the reconstruction method is based on linear operations, which benefits both physical interpretation and numerical implementation. Linear stochastic estimation (LSE) is a favourable mathematical tool for this purpose. Generally, LSE attempts to estimate the unknown variables based on the known variables by linear expressions containing empirical parameters. And the empirical parameters should be determined by minimizing the estimation errors in a statistical sense.

In wall-bounded turbulence community, the application of LSE in extracting conditionally-averaged structures has been widely reported, such as Adrian (1994), Christensen & Adrian (2001), and Wang *et al.* (2019). In these works, LSE served to estimate the neighbouring velocities of a reference even occurring at one spatial position, which is basically a point-based (zero dimension) estimation. As an extension, LSE based on one dimension (1D) signals has been reported by Tinney *et al.* (2006), who developed the so-called spectral LSE for this purpose. However, when it comes to the V2V reconstruction, the problem becomes estimating one 3D field (velocity field) based on another 3D field (vortex field), which is an unmitigated 3D-to-3D estimation and has not been reported in the turbulence community. In this section, the V2V reconstruction problem will be discussed under the general theoretical framework of LSE and the general governing equations will be deduced. Subsequently, the V2V reconstruction regarding wall-bounded turbulence is focused on, and the homogenous condition in the wall-parallel plane is employed to simplify the governing equations. At last, the corresponding implementation scheme and several considerations will be introduced in details.

For the convenience of the following discussion, we give the conventions for symbols first. Let $u$ and $\Lambda$ with a subscript $i$ or $j$ denote the $i$-th or the $j$-th component of velocity and vortex ($i=1,2,3$; $j=1,2,3$).

Einstein's summation convention is adopted so that two repeated indices occurring in one term trigger a summation for all possible realizations of these indices. $\mathbf{r} = (x, y, z)$, $\mathbf{r}' = (x', y', z')$ denote 3D spatial position vectors in the considered spatial domain $\Omega$.

*4.1 The general theory for V2V reconstruction*

In the context of LSE, the best estimation for $u_i$ based on $\Lambda_j$ should be a conditional average as $\langle u_i | \Lambda_j \rangle$ (Adrian 1994). In general, $\langle u_i | \Lambda_j \rangle$ could be estimated by optimizing the accuracy of estimation, which is denoted as $\hat{u}_i$ in this work. Suppose that $\hat{u}_i$ could be expressed by linear operators ($\mathcal{L}_{ij}$) acting on $\Lambda_j$, which yields

$$\hat{u}_i = \mathcal{L}_{ij} \Lambda_j, \tag{4.1}$$

where $[\mathcal{L}_{ij}]$ is a 3×3 matrix of linear operators. The linear operators can be viewed as continuous transformations on $\Lambda_j$, and no limitations on their realization forms are needed in this general analysis. (4.1) is the most general form for V2V reconstruction, and $\mathcal{L}_{ij}$ needs to be determined by minimizing the reconstruction deviations.

Considering both $\hat{u}_i$ and $u_i$ are 3D vector fields, the deviation ($D$) can be defined by an integral form as

$$D = \iiint_\Omega (\hat{u}_i - u_i)(\hat{u}_i - u_i) d\Omega \Big/ \iiint_\Omega d\Omega. \tag{4.2}$$

Facilitated by the above definition, the optimal $\mathcal{L}_{ij}$ can be determined by minimizing the deviation in a statistical sense, i.e.

$$\min \langle D \rangle, \tag{4.3}$$

where $\langle \cdot \rangle$ represents an ensemble average.

According to the calculus of variation (Appendix B), the above optimization problem is equivalent to the following equations,

$$\mathcal{L}_{ij} \langle \Lambda_j \Lambda_m(\mathbf{r}') \rangle = \langle u_i \Lambda_m(\mathbf{r}') \rangle \text{ for } \forall \Lambda_m(\mathbf{r}') \; (\mathbf{r}' \in \Omega; m = 1, 2, 3). \tag{4.4}$$

Equations from (4.4) constitute the *governing equations* for the linear operators $\mathcal{L}_{ij}$. Note that both $\Lambda_j$ and $u_i$ are 3D functions of the implicit position vector $\mathbf{r}$. And $\langle \Lambda_j \Lambda_m(\mathbf{r}') \rangle$ should also be regarded as a 3D function of $\mathbf{r}$ rather than $\mathbf{r}'$ when the linear operators ($\mathcal{L}_{ij}$) act on it. For any given $\Lambda_m(\mathbf{r}')$, the form of $\langle \Lambda_j \Lambda_m(\mathbf{r}') \rangle$ or $\langle u_i \Lambda_m(\mathbf{r}') \rangle$ indicates a conditional average for the event $\Lambda_m(\mathbf{r}')$. $\langle \Lambda_j \Lambda_m(\mathbf{r}') \rangle$ can be understood as the averaged vortex distribution given an event of $\Lambda_m$ occurring at $\mathbf{r}'$. And similarly, $\langle u_i \Lambda_m(\mathbf{r}') \rangle$ could be viewed as the corresponding velocity distribution. The occurrence of $\Lambda_m(\mathbf{r}')$ in (4.4) is much like putting a trial probe at $\mathbf{r}'$, which collects the statistical information about neighbouring vortex and velocity distributions. No matter where the probe ($\Lambda_m$) is put, the operators $\mathcal{L}_{ij}$ always act on the local statistical vortex distribution and return the corresponding velocity distribution.

Apparently, (4.4) is similar to the Yule-Walker equations for classical LSE (Adrian 1994). The later is typically employed to determine the parameters in the estimating expression, the number of which is usually small or finite at least. In this work, (4.4) is derived to determine the linear operators $\mathcal{L}_{ij}$, which have higher dimensions and could not be expressed by finite parameters in this discussion. To the best knowledge of the authors, it is the first time to report the governing equations regarding linear operators as (4.4) in the wall-bounded turbulence community. The starting point of the current method is to estimate one 3D field based on another given 3D field, which is a general problem, not only limited in wall-bounded turbulence. The estimation method based on (4.4) is called field-based linear stochastic estimation (FLSE) in the following discussion.

For a general case, the discretizing of (4.4) leads to a linear system with huge numbers of variables. For examples, if the spatial domain is discretized into $N_x \times N_y \times N_z$ computation nodes, numbers of the undetermined variables in $\mathcal{L}_{ij}$ would be $\left(3 \times N_x \times N_y \times N_z\right)^2$, which might be huge for a typical case. Fortunately, for wall-bounded turbulence, the flow shows homogeneous in the $x-y$ plane, which would significantly reduce the numbers of variables and will be discussed in the following subsection.

For the convenience of the following discussion, denote that

$$R_{jm}(\mathbf{r},\mathbf{r}') = \left\langle \Lambda_j \Lambda_m(\mathbf{r}') \right\rangle, \tag{4.5}$$

$$q_{im}(\mathbf{r},\mathbf{r}') = \left\langle u_i \Lambda_m(\mathbf{r}') \right\rangle. \tag{4.6}$$

Thus (4.4) can be rewritten as

$$\mathcal{L}_{ij} R_{jm}(\mathbf{r},\mathbf{r}') = q_{im}(\mathbf{r},\mathbf{r}') \quad \text{for} \quad \forall \ \mathbf{r}' \in \Omega; \ m = 1, 2, 3. \tag{4.7}$$

### 4.2 FLSE for wall-bounded turbulence

The estimated velocity $\hat{u}_i$ could be viewed as a result of the joint inducing effects caused by the neighbouring vortices. Suppose that the inducing effects of a vortex field could be expressed as an integral of the inducing effects of vortex vectors at different locations, which yields a form of integral transformation as

$$\hat{u}_i(\mathbf{r}) = \iiint_\Omega \varphi_{ij}(\mathbf{r};\mathbf{r}') \Lambda_j(\mathbf{r}') dx' dy' dz'. \tag{4.8}$$

In the equation, $\varphi_{ij}(\mathbf{r};\mathbf{r}')$ is a kernel function, which indicates the contribution of $\Lambda_j(\mathbf{r}')$ with unit magnitude to the $i$-th velocity component at the position $\mathbf{r}$. In fact, (4.8) provides a specific realization form for the linear operator $\mathcal{L}_{ij}$.

For wall-bounded turbulence, the flow could be treated as a homogenous field in the $x-y$ plane. Thus, the inducing effect of a vortex should keep invariable when its position is shifted in the $x-y$ plane, which yields

$$\varphi_{ij}(\mathbf{r};\mathbf{r}') = \varphi_{ij}(x,y,z;x',y',z') = \varphi_{ij}(x-x',y-y',z;0,0,z'). \tag{4.9}$$

For simplicity, it is denoted that

$$\varphi_{ij}(x-x',y-y',z;0,0,z') = \varphi_{ij}(x-x',y-y',z;z'). \tag{4.10}$$

By the way, such shifting-invariable property also holds for $R_{jm}(\mathbf{r},\mathbf{r}')$ and $q_{im}(\mathbf{r},\mathbf{r}')$, thus we also have $R_{ij}(x,y,z;z')$ and $q_{im}(x,y,z;z')$ as the simplifications for $R_{ij}(x,y,z;0,0,z')$ and $q_{im}(x,y,z;0,0,z')$.

Combining (4.8) and (4.9), we have

$$\hat{u}_i(\mathbf{r}) = \iiint_\Omega \varphi_{ij}(x-x', y-y', z; z')\Lambda_j(\mathbf{r}')dx'dy'dz'$$
$$= \int_{z'}\left(\iint_{x'-y'} \varphi_{ij}(x-x', y-y', z; z')\Lambda_j(\mathbf{r}')dx'dy'\right)dz' \quad . \tag{4.11}$$

The bracketed term is a 2D convolution operation in the $x-y$ plane, which could be denoted as a star. Thus, (4.11) can be simplified as

$$\hat{u}_i(\mathbf{r}) = \mathcal{L}_{ij}\Lambda_j = \int_{z'}\varphi_{ij}(\cdot,\cdot,z;z') * \Lambda_j(\cdot,\cdot,z')dz' , \tag{4.12}$$

where the position variable $x$ and $y$ is substituted as dots, indicating the convolution works in the $x-y$ plane. (4.12) is the formula for the so-called V2V reconstruction for wall-bounded turbulence, where $\varphi_{ij}(x,y,z;z')$ needs to be determined first. By definition, $\varphi_{ij}(x,y,z;z')$ reflects the inducing effect of individual vortices at different wall-normal positions, which is named as the inducing model function in this work.

Substituting the linear operators in (4.7) by the explicit expression of (4.12), the governing equations for $\varphi_{ij}(x,y,z;z')$ are obtained as

$$\int_{z''}\varphi_{ij}(\cdot,\cdot,z;z'') * R_{jm}(\cdot,\cdot,z'';z')dz'' = q_{im}(x,y,z;z') \text{ for } \forall z'; m=1,2,3. \tag{4.13}$$

Note that only the equations for $x'=0$ and $y'=0$ in (4.7) are collected into (4.13) since the number for unknown variables have been significantly reduced by using the homogenous condition.

## 4.3 Discretization and implementation for FLSE

Equations from (4.13) need to be solved in the discrete form. Suppose that the domain is discretized as $N_x \times N_y \times N_z$ computation nodes, and we use the symbols $a,b,c,c',c''$ as the indices for discretized variables of $x,y,z,z',z''$, respectively. The integral operation could be discretized as a summation implied by the repeated subscript of $c''$. Thus, we have

$$\varphi_{ij}(\cdot,\cdot,z_c;z''_{c''}) * R_{jm}(\cdot,\cdot,z''_{c''};z'_{c'}) = q_{im}(x_a,y_b,z_c;z'_{c'}) \text{ for } \forall c',m. \tag{4.14}$$

(4.14) is a linear system with underdetermined variables of $\varphi_{ij}(x_a,y_b,z_c;z''_{c''})$ for all the realizations of $i,j,a,b,c,c''$. Consistent with the former regulation, $x_a, y_b$ are omitted in (4.14) in order to indicate the working dimensions for the convolution operation. The regulation also avoids the occurrence of repeated $a,b$, which would cause confusion by erroneously triggering the automatic summation.

The number of underdetermined variables corresponding to all the possible realizations of $i,j,a,b,c,c''$ is $3 \times 3 \times N_x \times N_y \times N_z \times N_z$, which might be in the order of $10^9$. Fortunately, these variables do not need to be solved simultaneously. Closer observation on the indices occurring (4.14) reveals that $j,c''$ appears twice, which implies automatic summations and makes the corresponding variables coupled in the same equations. The variables corresponding to omitted indices $a,b$ are also coupled in the convolution operation. However, the indices $i$ and $c$ occurs only once, which implies that $\varphi_{ij}(x_a,y_b,z_c;z''_{c''})$ could be independently solved for the two indices. Specifically, for any prescribed indices $i$ and $c$, the equations from (4.14) constitute a closed linear system with respect to all the realizations of $\varphi_{ij}(x_a, y_b, z_c; z''_{c''})$ for $\forall j, a, b, c''$, which could be written as a matrix form.

$$\mathbf{R}\boldsymbol{\varphi}^{i,c} = \mathbf{q}^{i,c}, \tag{4.15}$$

where $\mathbf{R}$ is a matrix formed by arranging the realizations of $R_{jm}(x_a, y_b, z''_{c''}; z'_{c'})$; $\boldsymbol{\varphi}^{i,c}, \mathbf{q}^{i,c}$ are vectors formed by collecting all the realizations of $\varphi_{ij}(x_a, y_b, z_c; z''_{c''})$ and $q_{im}(x_a, y_b, z_c; z'_{c'})$ for prescribed indices $i$ and $c$. In this way, the equations from (4.14) could be divided into $N_z \times 3$ sets of reduced equations based on the different combinations of indices $i$ and $c$. Each set of equations contains $3 \times N_x \times N_y \times N_z$ variables and can be solved independently.

Besides directly solving (4.15), the problem could also be efficiently solved by using fast Fourier transform (FFT) technique. Perform 2D discrete FFT in the $x - y$ plane on (4.14), and apply the well-known convolution theorem of FFT. Let $\tilde{R}_{jm}, \tilde{\varphi}_{ij}$ denote the FFT results of $R_{jm}, \varphi_{ij}$, and $k_\alpha, k_\beta$ indicate the discrete wavenumbers along the $x, y$ direction. Thus we have

$$\sum_{c'',j} \tilde{R}_{jm}(k_\alpha, k_\beta, z''_{c''}; z'_{c'}) \tilde{\varphi}_{ij}(k_\alpha, k_\beta, z_c; z''_{c''}) = \tilde{q}_{im}(k_\alpha, k_\beta, z_c; z'_{c'}) \quad \text{for} \quad \forall c', m. \tag{4.16}$$

Einstein's summation convention is abandoned in this equation since the repeated indices $\alpha, \beta$ do not imply a summation here. For numerical implementation, (4.16) should also be written as the matrix form. In this case, we can see $\tilde{\varphi}_{ij}(k_\alpha, k_\beta, z_c; z''_{c''})$ is independent with the indices of $i, \alpha, \beta, c$. Therefore, for fixed indices of $i, \alpha, \beta, c$, we can collect all realizations for $\tilde{R}_{jm}(k_\alpha, k_\beta, z''_{c''}; z'_{c'})$ and arrange them into a matrix ($\mathbf{A}^{\alpha,\beta}$). Similarly, $\tilde{\varphi}_{ij}$ and $\tilde{q}_{im}$ are arranged as vector forms as $\boldsymbol{\varphi}^{i,\alpha,\beta,c}$ and $\mathbf{b}^{i,\alpha,\beta,c}$ for prescribed indices $i, \alpha, \beta, c$.

Facilitated by these conventions, (4.16) could be rewritten as

$$\mathbf{A}^{\alpha,\beta} \boldsymbol{\varphi}^{i,\alpha,\beta,c} = \mathbf{b}^{i,\alpha,\beta,c}. \tag{4.17}$$

For each set of indices $i, \alpha, \beta, c$, the equations of (4.17) constitute a closed linear system for $\boldsymbol{\varphi}^{i,\alpha,\beta,c}$, which contains $N_z \times 3$ variables. After all the equations are solved, $\tilde{\varphi}_{ij}(k_\alpha, k_\beta, z_c; z''_{c''})$ is recovered based on all the results. Inverse FFT (IFFT) operation is subsequently performed on $\tilde{\varphi}_{ij}(k_\alpha, k_\beta, z_c; z''_{c''})$ to return $\varphi_{ij}(x_a, y_b, z_c; z''_{c''})$. For clarify, a procedure card illustrating the FFT implementation has been provided in figure 7. This card also incorporates the strategy of dealing with the numerical instability issue, which will be introduced in the next subsection.

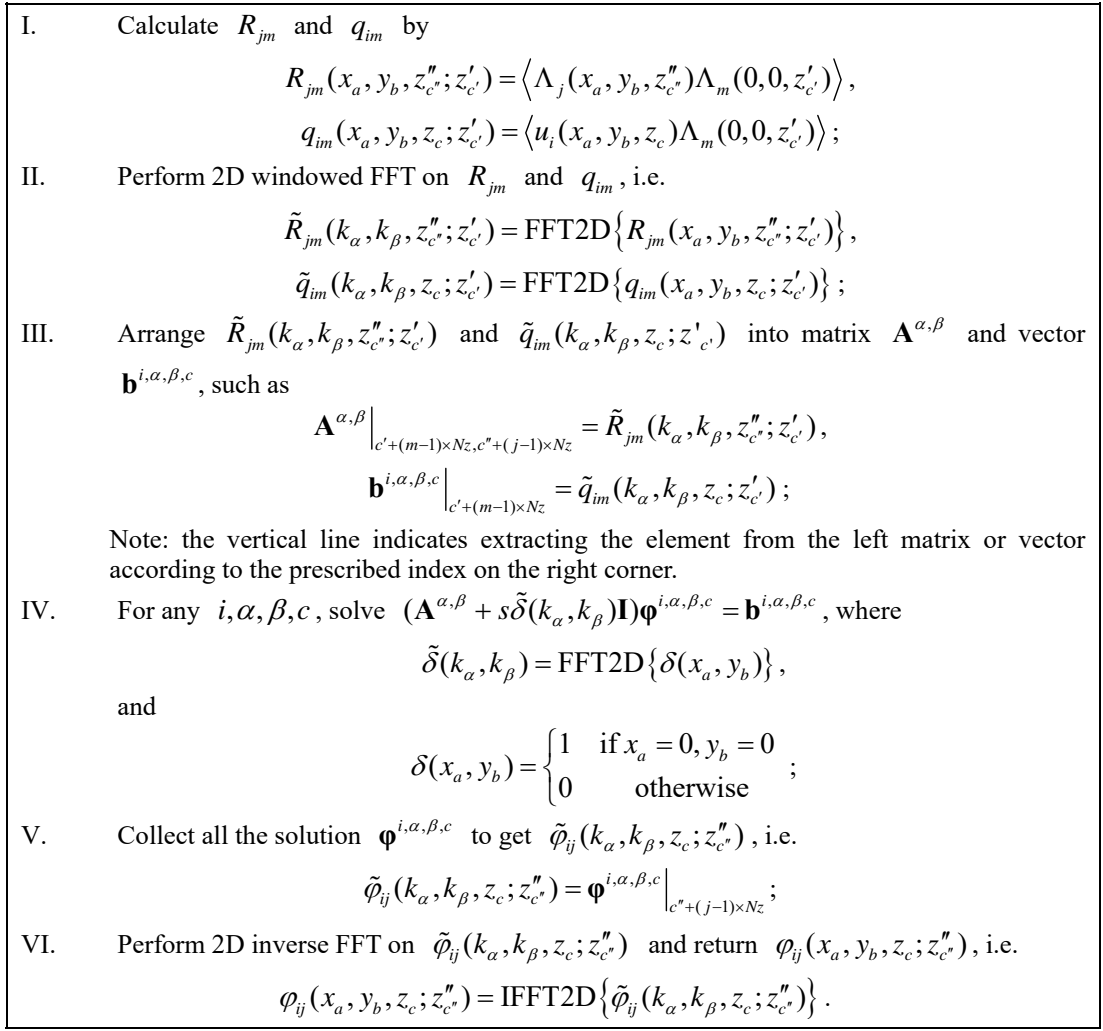

| | |
|---|---|
| I. | Calculate $R_{jm}$ and $q_{im}$ by $$R_{jm}(x_a, y_b, z''_{c''}; z'_{c'}) = \langle \Lambda_j(x_a, y_b, z''_{c''}) \Lambda_m(0,0,z'_{c'}) \rangle,$$ $$q_{im}(x_a, y_b, z_c; z'_{c'}) = \langle u_i(x_a, y_b, z_c) \Lambda_m(0,0,z'_{c'}) \rangle;$$ |
| II. | Perform 2D windowed FFT on $R_{jm}$ and $q_{im}$, i.e. $$\tilde{R}_{jm}(k_\alpha, k_\beta, z''_{c''}; z'_{c'}) = \text{FFT2D}\{R_{jm}(x_a, y_b, z''_{c''}; z'_{c'})\},$$ $$\tilde{q}_{im}(k_\alpha, k_\beta, z_c; z'_{c'}) = \text{FFT2D}\{q_{im}(x_a, y_b, z_c; z'_{c'})\};$$ |
| III. | Arrange $\tilde{R}_{jm}(k_\alpha, k_\beta, z''_{c''}; z'_{c'})$ and $\tilde{q}_{im}(k_\alpha, k_\beta, z_c; z'_{c'})$ into matrix $\mathbf{A}^{\alpha,\beta}$ and vector $\mathbf{b}^{i,\alpha,\beta,c}$, such as $$\mathbf{A}^{\alpha,\beta}\big|_{c'+(m-1)\times Nz, c''+(j-1)\times Nz} = \tilde{R}_{jm}(k_\alpha, k_\beta, z''_{c''}; z'_{c'}),$$ $$\mathbf{b}^{i,\alpha,\beta,c}\big|_{c'+(m-1)\times Nz} = \tilde{q}_{im}(k_\alpha, k_\beta, z_c; z'_{c'});$$ Note: the vertical line indicates extracting the element from the left matrix or vector according to the prescribed index on the right corner. |
| IV. | For any $i, \alpha, \beta, c$, solve $(\mathbf{A}^{\alpha,\beta} + s\tilde{\delta}(k_\alpha, k_\beta)\mathbf{I})\boldsymbol{\varphi}^{i,\alpha,\beta,c} = \mathbf{b}^{i,\alpha,\beta,c}$, where $$\tilde{\delta}(k_\alpha, k_\beta) = \text{FFT2D}\{\delta(x_a, y_b)\},$$ and $$\delta(x_a, y_b) = \begin{cases} 1 & \text{if } x_a = 0, y_b = 0 \\ 0 & \text{otherwise} \end{cases};$$ |
| V. | Collect all the solution $\boldsymbol{\varphi}^{i,\alpha,\beta,c}$ to get $\tilde{\varphi}_{ij}(k_\alpha, k_\beta, z_c; z''_{c''})$, i.e. $$\tilde{\varphi}_{ij}(k_\alpha, k_\beta, z_c; z''_{c''}) = \boldsymbol{\varphi}^{i,\alpha,\beta,c}\big|_{c''+(j-1)\times Nz};$$ |
| VI. | Perform 2D inverse FFT on $\tilde{\varphi}_{ij}(k_\alpha, k_\beta, z_c; z''_{c''})$ and return $\varphi_{ij}(x_a, y_b, z_c; z''_{c''})$, i.e. $$\varphi_{ij}(x_a, y_b, z_c; z''_{c''}) = \text{IFFT2D}\{\tilde{\varphi}_{ij}(k_\alpha, k_\beta, z_c; z''_{c''})\}.$$ |

FIGURE 7. A procedure card for the FFT implementation of FLSE.

### 4.4 Issues and considerations for numerical implementation

Although the FFT implementation significantly improves the computation efficiency, some considerations need to be addressed. In theory, the convolution theorem for FFT exactly holds for the circular convolution operation, which imposes a periodic extension on the input functions. If the computation domain $\Omega$ is not large enough, magnitudes of $R_{jm}$ or $q_{im}$ would not decay sufficiently at borders. The periodical extension leads to strong disconnections on borders, which would cause numerical instability in the FFT results. Therefore, windowed FFT technique is suggested to deal with this issue. In this case, FFT method would be viewed as an approximate method, whose accuracy depends on the computation domain and needs to be further validated by the estimating results.

As we can see from the definitions, both $\mathbf{R}$ and $\mathbf{A}^{\alpha,\beta}$ should be symmetric and non-negative defined. However, the numerical stability for solving (4.15) or (4.17) requires that $\mathbf{R}$ and $\mathbf{A}^{\alpha,\beta}$ should be sufficiently positive definite, which mean all the eigenvalues should be large than a positive critical number. In practical implementation, both $\mathbf{R}$ and $\mathbf{A}^{\alpha,\beta}$ have zeros or very small eigenvalues, which cause ill-posed systems. These eigenvalues with zero or small magnitude are caused by the redundancy of the system matrices. Recalling the discussion in Section 3, the vortex fields approximately satisfy the continuity equation, which means that the three vortex components are not wholly independent, and information redundancy exists in their correlation matrices ( $\mathbf{R}$ and $\mathbf{A}^{\alpha,\beta}$ ). Such implicit information redundancy in system matrices would cause multiple feasible solutions. In order to avoid this problem, Tikhonov regularization (Kress 2014) is suggested to improve the numerical stability, which reshapes the original (4.15) and (4.17) into the following forms, respectively.

$$(\mathbf{R} + s\mathbf{I})\boldsymbol{\varphi}^{i,c} = \mathbf{q}^{i,c}, \tag{4.18}$$

$$\left(\mathbf{A}^{\alpha,\beta} + s\tilde{\delta}_c(k_\alpha,k_\beta)\mathbf{I}\right)\boldsymbol{\varphi}^{i,\alpha,\beta,c} = \mathbf{b}^{i,\alpha,\beta,c}, \tag{4.19}$$

where $\mathbf{I}$ is the identity matrix and $s$ is the regularization parameter. The summation convention is cancelled in the equation. $\tilde{\delta}(k_\alpha,k_\beta)$ is the FFT result for a discrete 2D Dirac function $\delta(x_a,y_b)$, which is defined as

$$\delta(x_a,y_b) = \begin{cases} 1 & \text{for } x_a = 0, y_b = 0 \\ 0 & \text{otherwise} \end{cases}. \tag{4.20}$$

In (4.18) and (4.19), $s$ plays a role in improving the statistical convergence and controlling the smoothness of the solution, which will be tested in Section 5.

The FFT implementation of FLSE is a reminiscence for spectral linear stochastic estimation (SLSE) proposed by Tinney *et al.* (2006). Baars, Hutchins & Marusic (2016) applied SLSE to predict the velocity signals in wall-bounded turbulence and refined the inner-outer interaction model proposed by Marusic, Mathis & Hutchins (2010). More recently, Baars & Marusic (2020) developed a data-driven decomposition scheme to decouple inner-outer coherence based on SLSE. Basically, SLSE reconstructs the prediction for 1D signal by minimizing the estimation error in spectral space, which avoids the phase mismatch problem. While expanding SLSE from 1D to 2D in *x-y* plane is straightforward, further expanding to 3D application needs specially dealing with the wall-normal dimension. In this work, FLSE offers a general routine to estimate one field based on another, regardless of the dimensions. As the implementation of FLSE, the FFT method expands spectra-based LSE to 3D application in wall-bounded turbulence. The expansion brings new concerns about the accuracy and numerical instability issues, which explains why the windowed FFT scheme and the Tikhonov regularization are not involved in SLSE.

## 5. Numerical validation and application of FLSE

This section will focus on the validation and application of FLSE. The correlation functions and the solved inducing model functions will be displayed, which helps to illustrate how FLSE works, and also benefits the discussion about how to choose the statistical range and the regularization parameters. The performance of FLSE will be comprehensively compared with Biot-Savart law with the aim of promoting the FLSE method. The reconstruction accuracy of both methods applied at different wall-normal positions and dealing with different scales of motions will be investigated. Subsequently, the performance of FLSE in recovering energy spectra is accessed, which provides some reference for the vortex-based modelling works. A novel application of FLSE in shedding light on the energy contributions from different heights will be introduced, and the results will be linked to the topic of the inner-outer interaction. At last, the high-order moments for FLSE-reconstructed fields will be evaluated, which will arouse the discussions about the limitations of FLSE.

## 5.1 Correlation functions: $R_{jm}$ and $q_{im}$

Before FLSE, the self-correlation function for vortex fields ($R_{jm}$) and vortex-velocity correlation function ($q_{im}$) need to be extracted from data. In this work, these correlation functions are calculated based on twelve large-size instantaneous DNS fields (as shown in table 1). The ensemble average is implemented by averaging on the statistically uniform dimensions, including the $x-y$ plane and time dimension. Thus, the two relation functions in this implementation are calculated by

$$R_{jm}(x,y,z;z') = \left\langle \Lambda_j(x_0+x, y_0+y, z)\Lambda_j(x_0, y_0, z') \right\rangle_{x_0, y_0, t}, \tag{5.1}$$

$$q_{im}(x,y,z;z') = \left\langle u_i(x_0+x, y_0+y, z)\Lambda_m(x_0, y_0, z') \right\rangle_{x_0, y_0, t}. \tag{5.2}$$

Noting that $\langle \cdot \rangle_{x_0, y_0, t}$ denotes averaging on the variables of $x_0, y_0, t$.

In order to avoid the resolution issue, the calculation is based on the original computational grid of DNS. The statistical range for correlation functions is limited as $x^+ \in [-0.5\delta^+, 0.5\delta^+]$, $y^+ \in [-0.25\delta^+, 0.25\delta^+]$, $z^+ \in [5, 0.3\delta^+]$, which results in a computation grid of 179×155×91. The investigation of del Alamo et al. (2006) on size of attaching clusters supported the aspect ratio prescribed in this work. The wall-normal range covers the whole buffer layer and logarithmic layer, which contain most of the vortex populations in TBL. It should be pointed out that the statistical range for correlation functions also determines the calculation domain of $\varphi_{ij}$, which is correlated with reconstruction accuracy. More arguments for choosing this statistical range will be emphasized in the following discussion.

Figure 8 displays contours of $R_{11}$ and $q_{11}$ as functions of $x$ and $y$ for three wall-normal positions ($z'^+ = z^+ = 15.4, 119.4, 359.3$). At first glance, these contours are reasonably smooth and symmetric, indicating that the statistical results are well converged. At $z^+ = 15.4$, multi-streak patterns occur around the centre points for both $R_{11}$ and $q_{11}$, which are imprints of the quasi-streamwise vortices and low-speed streaks populated in this region. When $z^+$ increase from 119.4 to 359.3, both $R_{11}$ and $q_{11}$ display a self-similar increase in size, supporting the famous attached eddy hypothesis. Distinctive characteristics for $R_{11}$ and $q_{11}$ are also observed: while $R_{11}$ remains a sharp peak around the central point, $q_{11}$ shows two flat peaks filling a large portion of the area in the plots. At the edge of the wall-normal range considered ($z^+ = 359.3$), bulks of the two $q_{11}$ peaks still remain in the statistical range although they appear to be truncated at boundaries to some extent. A much larger statistical range is needed if one attempts to cover the whole pattern for $q_{11}$ in the $x-y$ plane at $z^+ = 359.3$, which poses a severe challenge for both the amount of statistical data required and also numerical implementation. The statistical range employed in this work is based on the comprehensive considerations on the computation efficiency we can bear and the reconstruction accuracy we have expected. In fact, most of the results in this work will focus on the regions below the logarithmic layer, particularly for $z^+ < 120$, where the correlation peaks for $R_{11}$ is satisfyingly covered in the statistical range. More importantly, the resulting reconstruction accuracy in this region is good enough to promote the FLSE method and to draw some revealing conclusions as we can see in the following discussions.

Once $R_{jm}$ and $q_{im}$ for $i, j, m = 1, 2, 3$ are obtained, the vortex model function $\varphi_{ij}(x, y, z; z')$ can be solved based on (4.18) or (4.19). Considering the computation nodes involved are quite large (about 2.5 million) in this case, the FFT implementation scheme is employed based on the processing procedures shown in figure 7. To avoid the interruptions at boundaries, both $R_{11}$ and $q_{11}$ are weighted by a Hanning window before the FFT operation. Weighting a signal by a Hanning window procedure will force the input signal to decay to zeros at boundaries, which benefits the numerical stability. Testing on the Hanning window and the regularization parameter will be introduced in the following subsection.

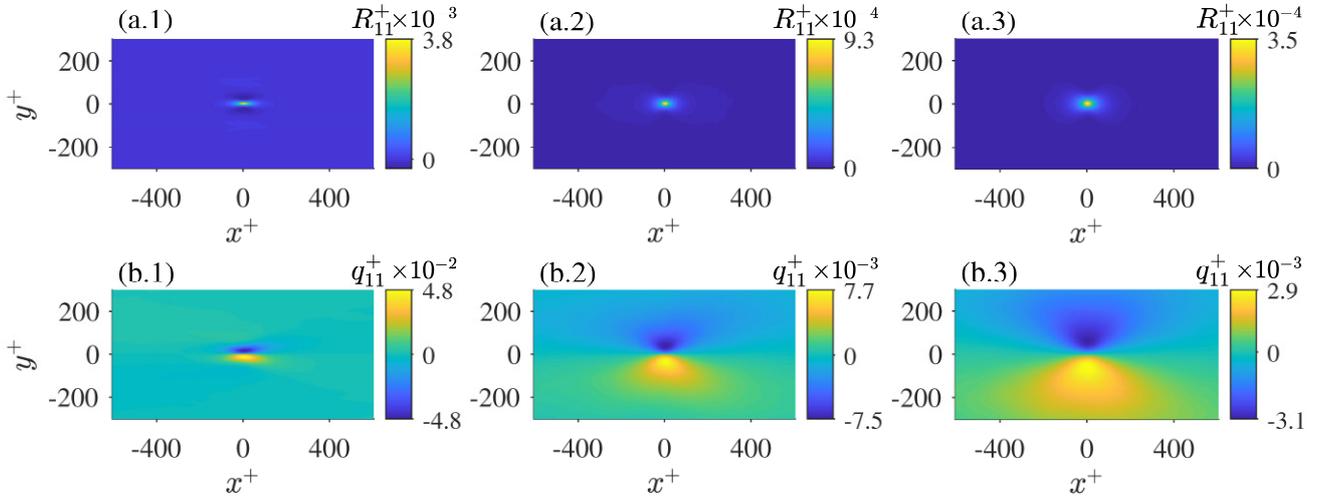

FIGURE 8. Contours of $R_{11}^+(x, y, z; z')$ (the first row) and $q_{11}^+(x, y, z; z')$ (the second row). From left to right, the contours correspond to $z^+ = z'^+ = 15.4$, $z^+ = z'^+ = 119.4$, and $z^+ = z'^+ = 359.3$, respectively.

*5.2 Regularization parameter and the Hanning window*

The performance of FLSE is determined by the inducing model functions. Figure 11 shows $\varphi_{13}(x, y, z; z)$ in the $x - y$ plane, which reflects the in-plane inducing effect of the individual wall-normal vortex on the streamwise velocity component. Figure 9 displays the results from three calculation cases: two cases passing the typical processing routines with $s = 0.82 \times 10^{-6} \mu_{max}$ and $s = 0.82 \times 10^{-4} \mu_{max}$ respectively, and one case skipping window-weighting before FFT with $s = 0.82 \times 10^{-4} \mu_{max}$. Herein, $\mu_{max}$ is the maximum eigenvalue for all the realizations of $\mathbf{A}^{a,b}$. In order to make it clear, only a small region with a streamwise range of $\pm 150$ and a spanwise range of $\pm 75$ is displayed. It shows that while figure 9(a) and figure 9(b) show similar contours, the contour maps of figure 9(c) suffer from numerical issues and present spurious patterns. It indicates that the Hanning window is necessary for this FFT implementation of FLSE in order to obtain credible results. In figure 9(a) or figure 9(b), $\varphi_{13}$ shows two streamwise-extending streaks at $z^+ = 15.4$, which reflects the basic flow feature in this region. At $z^+ = 119.4$, the streak patterns shrink along the streamwise direction and form a velocity distribution resembling an intense swirl in the $x - y$ plane. The wall-normal variation of the inducing model function gives signs of adaption to local flow features, which is vital for the success of V2V reconstruction. By comparing figure 9(a) and figure 9(b), it can be found that adjusting the value of $s$ works as a dilation or erosion effect on the $\varphi_{13}$ pattern. Larger $s$ helps to obtain smoother and more symmetrical contour patterns, which implies a good convergence. For a very large $s$, the equation system of (4.19) will result in a solution very close to $\mathbf{b}^{a,b,c}$ (i.e. $q_{im}$), which shows a quite flat pattern in the $x-y$ plane as shown in figure 8. At last, although $\varphi_{13}$ from case 1 and case 2 shows a clear difference in the magnitude, the corresponding reconstructed velocity fields are quite similar. Quantitative examination shows that the averaged CC for the reconstructed velocity fields based on the two parameters is about 0.995 for three wall-normal positions considered, which validates the robust performance of FLSE.

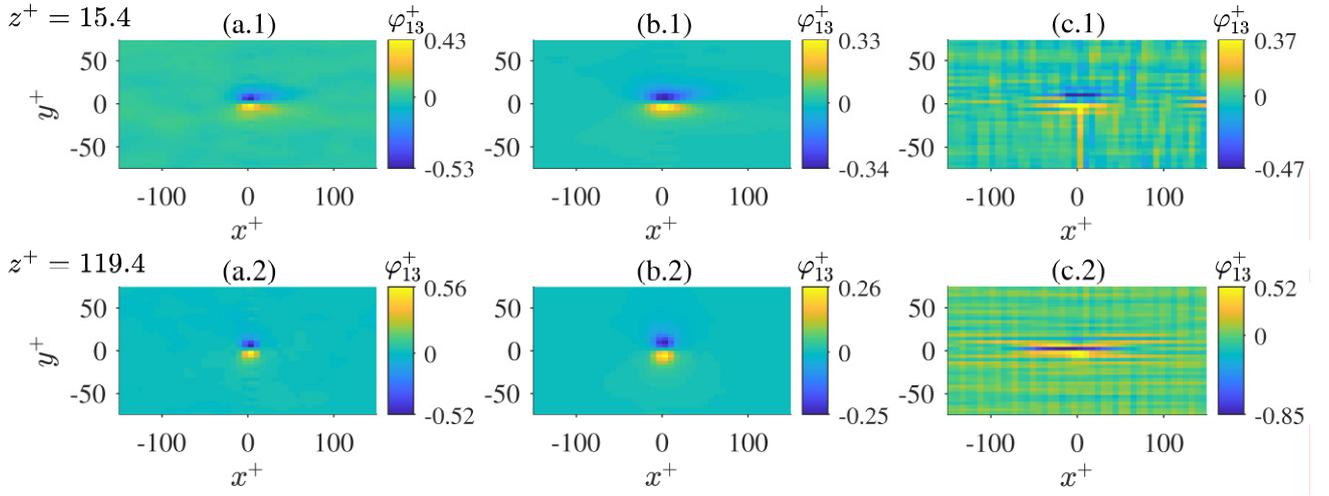

FIGURE 9. $\varphi_{13}^+(x,y,z;z')$ as a function of $x^+$ and $y^+$ for $z^+ = z'^+ = 15.4$ (the first row) and $z^+ = z'^+ = 119.4$ (the second row). The left, middle and right columns correspond to three calculation cases. (a): Calculated by a typical routine with $s = 0.82 \times 10^{-6} \mu_{max}$; (b): calculated by a typical routine with $s = 0.82 \times 10^{-4} \mu_{max}$; (c): calculated by a routine neglecting the window-weighting procedure with $s = 0.82 \times 10^{-4} \mu_{max}$.

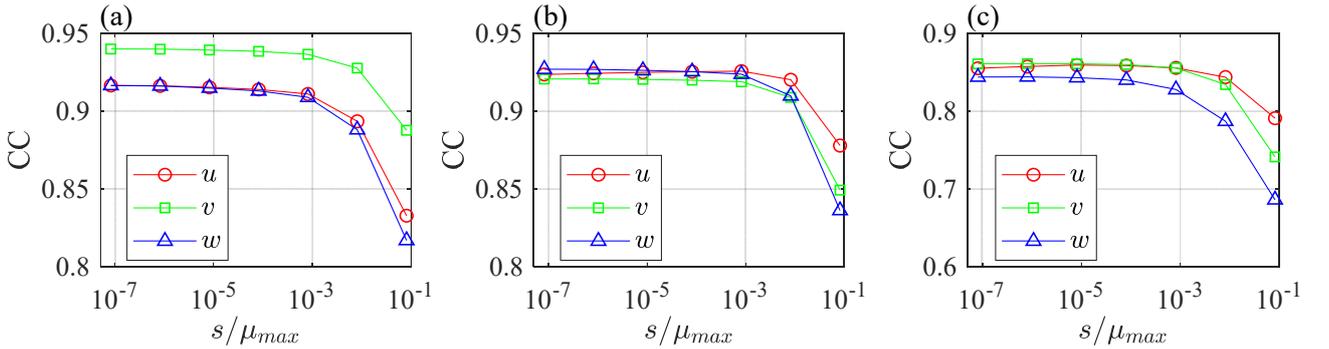

FIGURE 10. Correlation coefficients for reconstructed/original velocity fields at $z^+ = 15.4, 119.4, 359.3$ (a-c, respectively) as functions of the regularization parameter.

To quantitatively investigate the influence of $s$ in the reconstruction accuracy, a series of $s$ ranging from $s = 0.82 \times 10^{-7} \mu_{max}$ to $0.82 \times 10^{-1} \mu_{max}$ (logarithmically spaced) are trialled. The minimum $s$ employed in this test is very close to the lower limit for keeping the numerical stability. The solved inducing model functions from (4.19) are used to reconstruct a large DNS field (as detailed by table 1) based on (4.12). CCs for reconstructed velocity fields and original velocity fields at three wall-normal positions are displayed in Fig 10 as functions of $s$. The results show that smaller $s$ corresponds to higher reconstruction accuracy, although the results for $s/\mu_{max} < 10^{-3}$ are quite close. The best performance of FLSE achieves a CC of larger than 0.9 for all the three velocity components in figure 10(a) and figure 10(b), which indicates that the reconstructed velocity field is very similar to the real one. For figure 10(c), the reconstruction accuracy is lower since it corresponds to the edge of reconstructing domain, and only the vortices below this position are considered in the reconstruction. Based on the discussions for figure 9 and figure 10, this work will adopt $s = 0.82 \times 10^{-4} \mu_{max}$ in the following analysis, which corresponds to a good performance in both convergence and accuracy. It is worthy to point out that the reconstruction results for $s/\mu_{max} < 10^{-3}$ are quite close to each other, and thus prescribing any other $s$ in this range brings no substantial change in the following results of this work.

### 5.3 Comparing with Biot-Savart law

Biot-Savart (BS) law provides an explicit formula to reconstructing velocity fields based on vorticity fields. In theory, the law strictly holds for the situations where vorticities reside in a limited domain without walls. Several investigators (Perry & Marusic 1995; Marusic 2001; de Silva et al. 2016a) have employed BS law to reconstruct wall-bounded turbulence based on the random arrangement of typical vortex tubes. According to the inherent correlation between vorticity and vortex, the BS law could be employed to reconstruct the velocity field directly based on the vortex field. The BS law with respect to the vortex-based reconstruction can be expressed as

$$\mathbf{u}(\mathbf{r}) = \frac{1}{4\pi} \int_\Omega \frac{\mathbf{\Lambda}(\mathbf{r}') \times (\mathbf{r} - \mathbf{r}')}{|\mathbf{r} - \mathbf{r}'|^3} d\Omega, \qquad (5.3)$$

In (5.3), $\mathbf{r}'$ is the running position vector for the volume integration and $\mathbf{r}$ is the reference position vector. For convenience in numerical implementation, the equation is reformed as

$$u_i(\mathbf{r}) = \int_\Omega P_{ij}(\mathbf{r} - \mathbf{r}') \Lambda_j(\mathbf{r}') d\Omega = P_{ij} * \Lambda_j, \qquad (5.4)$$

where $*$ indicates a 3D convolution operation, different from the 2D convolution in (4.12). And the convolution kernel $P_{ij}$ is determined by

$$[P_{ij}] = \frac{1}{4\pi(x^2 + y^2 + z^2)^{3/2} + \varepsilon} \begin{bmatrix} 0 & z & -y \\ -z & 0 & x \\ y & -x & 0 \end{bmatrix}. \qquad (5.5)$$

where $\varepsilon$ is a small parameter, employed to avoid the singularity at the origin. In this work, we prescribe $\varepsilon^+ = 10^{-8}$, which makes $P_{ij}(0,0,0) = 0$ and brings only a tiny change (in the order of $10^{-8}$) on the values of $P_{ij}$ at other positions. The convolutional kernel of BS law ($P_{ij}$) resembles the inducing model function $\varphi_{ij}$ in FLSE. Differently, BS law employs a fixed and isotropic model function while FLSE adopts a data-driven inducing model, which involves the statistical imprints of flows. For implementation, $P_{ij}$ is truncated and discretized on the same grid as $\varphi_{ij}$. In order to make the reconstruction results of BS law quantitatively close to the truth velocity fields, the input vortex field should be multiplied by a constant to make the magnitude comparable to the real vorticity magnitude. Pirozzoli *et al.* (2010) employed a multiplier of two when dealing with this issue, which was derived based on the situation of a rigidity rotation. In this work, the multiplier is determined by minimizing the mean squares of the deviations between reconstructed results and original DNS results.

The reconstructed $u$ fields based on both BS law and FLSE are displayed in figure 11. The $u$ field displayed in the figure covers an area of $4\delta \times 2\delta$ in the streamwise-spanwise plane, which is four times as large as the calculation domain of $\varphi_{ij}$. Results show that FLSE performs surprisingly well in both the near-wall region and the logarithmic region. The turbulent motions with scales ranging from the spacing of low-speed streaks to the boundary thickness are well recovered. As emphasized in the foregoing discussion, vortices are very sparse in space, and the vortex field contains almost as less information as a scalar field, which poses great challenges for the reconstruction. The success in recovering velocity fields based on vortex structure should be owing to the inherent coherence of vortex structures which could be well modelled as empirical functions as $\varphi_{ij}$. BS law performs as well as FLSE for the logarithmic region but fails to reconstruct the low-speed streaks in the near-wall region. The lousy performance of BS law in the near-wall region is expectable since the distributions of vortices and vorticities are quite different below the buffer layer, as shown in figure 1. In fact, the wall effect is not considered in the derivation of the BS law, which indicates that even vorticity-based reconstruction is not valid in the region very close to the wall, let alone the vortex-based reconstruction. The discussion reminds us that BS law cannot be employed below the logarithmic region for the V2V reconstruction.

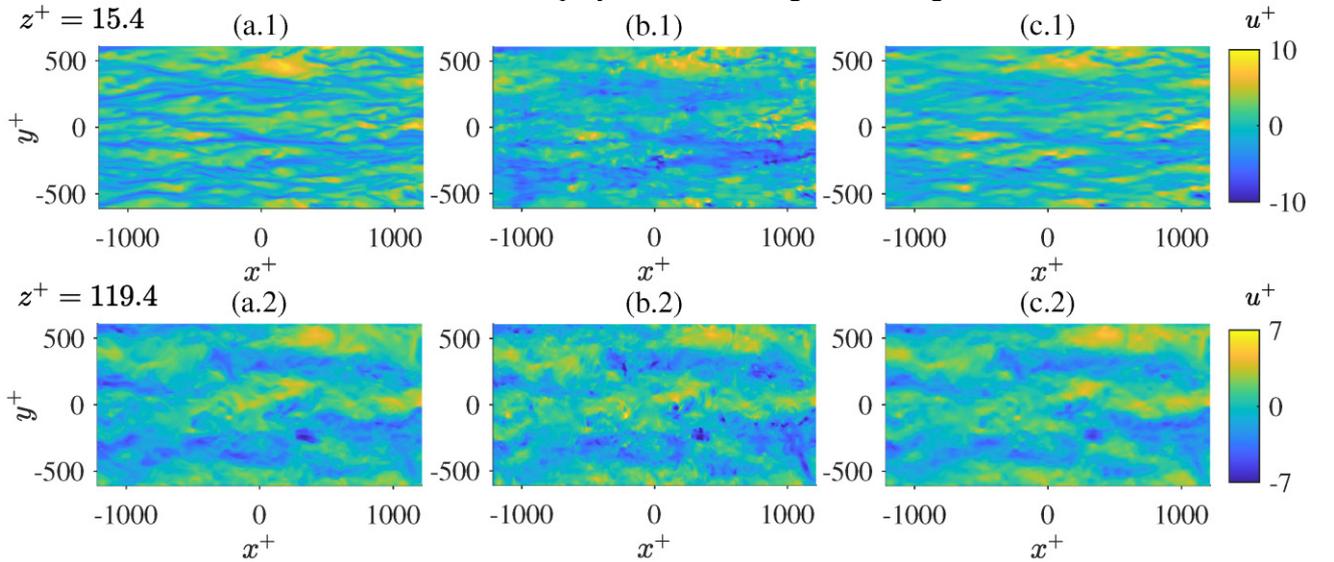

FIGURE 11. Instantaneous $u^+$ fields from $z^+ = 15.4$ (the first row) and $z^+ = 119.4$ (the second row). (a) Original DNS data, (b) reconstructed $u^+$ based on BS law, (c) reconstructed $u^+$ based on FLSE.

Quantitative comparisons between the performances of FLSE and BS law are provided in figure 12. Figure 12(a) shows the CCs for original/reconstructed $u, v, w$ as functions of $z$. The most attractive point indicated from this plot is the good performance of FLSE for $z^+ < 100$. In this region, the CCs from FLSE remain a steady variation above 0.9 while the CCs from the BS law encounter a dramatic drop for $u, v, w$. In the logarithmic layer, both FLSE and BS law perform well, achieving CCs of larger than 0.9 for $u, v, w$. BS law perform slightly better than LSE for $z^+ > 150$, which supports the application of BS law in the logarithmic layer (Perry & Marusic 1995; Marusic 2001; de Silva *et al.* 2016a) and makes BS law very competitive in this region considering its simplicity in implementation. The imperfect performance of FLSE for $z^+ > 150$ is attributed to the limited calculation domain for $\varphi_{ij}$, which would be further analyzed in the following scale-specific assessment. At last, the performances of both methods deteriorate at the upper boundary of the reconstruction domain, consistent with the results of figure 10.

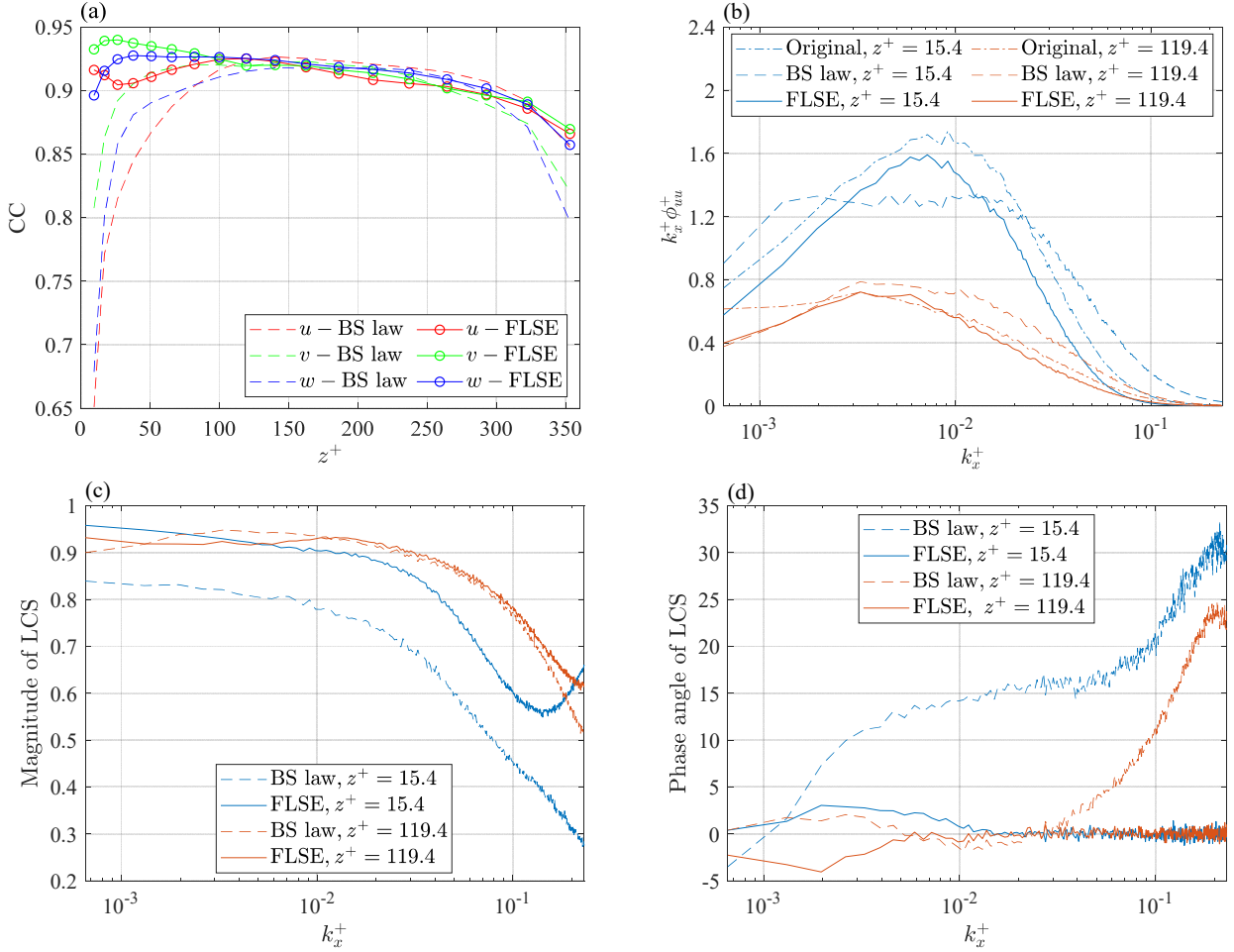

FIGURE 12. Quantitative comparisons between the reconstruction results of FLSE and BS law. (a) Correlation coefficients for original/reconstructed fields as functions of $z^+$; (b) premultiplied streamwise energy spectra of $u^+$ from the original DNS data and the reconstructed results as functions of $k_x^+$; (c) magnitudes of streamwise LCS for original/reconstructed $u^+$; (d) phase angles of streamwise LCS for original/reconstructed $u^+$.

The premultiplied spectra shown in figure 12(b) provide scale-by-scale comparisons between the reconstructed results and original DNS data. Only the streamwise spectra for $u$ component (denoted as $\phi_{uu}$) are taken as examples in this discussion. At $z^+ = 15.4$, the spectra from BS law show a plateau at $k_x^+ = 1.5 \times 10^{-3} \sim 2 \times 10^{-2}$ ($k_x^+$ is the streamwise wavenumber), which is obviously different from the single peak pattern of the real spectra. The spectra from FLSE show slightly lower energy for all the wavenumbers, but the shape of the spectral curve resembles the real spectra. At $z^+ = 119.4$, the spectra from FLSE are very close to the real spectra except for the very low wavenumbers, where the spectra from both FLSE and BS law are lower than the real spectra. For high wavenumbers, the spectra from FLSE tend to be lower than the real spectra while the spectra from BS law tend to be higher to more extent. Totally, the results show that FLSE performs better in recovering the energy spectra.

Generally, two signals sharing the same spectra might be completely different since the spectra are integral quantities, which are robust to local events. To further evaluate the reconstruction, the LCSs for reconstructed and original $u$ along the streamwise direction are calculated. As introduced in Section 2.3, the magnitude and phase angle of LCS indicate the scale-by-scale correlation coefficient and scale-by-scale phase deviation between two input signals, respectively. In the best case, if the magnitude of LCS equals one and the phase angle equals zeros for each scale, it can be inferred that the two input signals can be exactly matched by multiplying a constant. LCS offers a strict evaluating indicator for the performance of FLSE in reconstructing turbulent motions with different scales. One particular scale deserving more attention is the streamwise length of the calculation domain for $\varphi_{ij}$ (denoted as $L_x$), which corresponds to $k_x^+ \approx 5 \times 10^{-3}$.

The magnitudes and phase angles of LCS are displayed in figure 12(c-d). Plenty of information can be drawn from these two plots. Firstly, the overall performance and the effective scale range for FLSE are focused. Figure 12(c) shows that for both $z^+ = 15.4$ and $z^+ = 119.4$ the magnitudes of LCS for FLSE-reconstructed/original $u$ keep larger than 0.9 for the scale range of $k_x^+ < 2 \times 10^{-2}$, which corresponds to a bulk of the total energy according to figure 12(b). The good performance for FLSE is also observed from the phase angles of LCS, which fluctuate in a small range of ±5 degrees for the whole wavenumber range considered. The most striking point is that FLSE performs well even for the largest scale considered, which is eight times as large as $L_x$. Another interesting point is that mostly the LCS magnitude decreases with the increase of the wavenumbers, which indicates that the large-scale motions are easier to be reconstructed than the small-scale fluctuations. This trend is easy to explain since the large structures are composed of large numbers of independent vortex tubes. While the individual vortex tubes suffer from strong random fluctuations in the logarithmic layer, their joint inducing effect is more robust due to the offset effect of independent random fluctuations. This explanation needs to be revised for the case of $z^+ = 15.4$ since the background fluctuation is comparatively weak in this region and the viscous effect needs to be considered.

On the other hand, figure 12(c-d) also allow a further comparison between FLSE and BS law. For the results of $z^+ = 15.4$, the advantages of FLSE are observed from both figure 12(c) and figure 12(d). FLSE achieves a larger LCS magnitude and smaller LCS angles for almost all the wavenumbers. For the results of $z^+ = 119.4$, BS law has slightly better performance for $k_x^+ < 10^{-2}$, which corresponds to a length scale of about $0.5L_x$. A similar demarcation with $k_x^+ \approx 10^{-2}$ is also observed in the performance of FLSE in figure 12(d). For $k_x^+ > 10^{-2}$, the phase angle for LCS fluctuates around zero; yet for $k_x^+ < 10^{-2}$ the phase angle deviate from zero to a negative or positive value. The explanation is clear since the governing equations for $\varphi_{ij}$ are limited in the calculation domain. The intention of minimizing the reconstruction deviations is not forced for the scales larger than $L_x$. These results constitute an all-round assessment on the reconstruction performance of FLSE and BS law, which benefits the application of both methods. Mostly, FLSE achieves better reconstruction accuracy, particularly for the buffer layer, which will be focused in the following analysis.

*5.4 Full energy spectra of FLSE-reconstructed TBL*

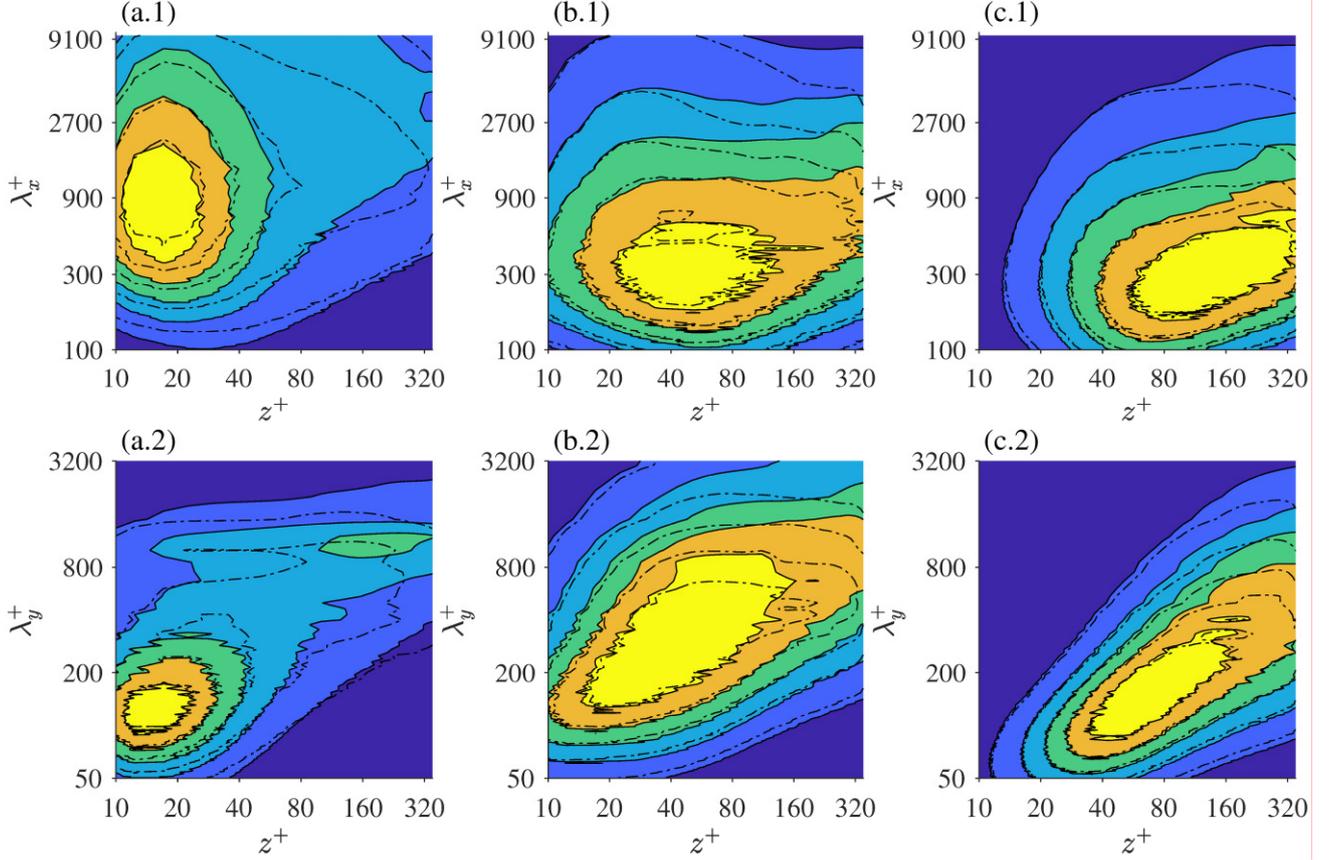

FIGURE 13. Contour maps showing the variation of premultiplied spectra for $u^+, v^+, w^+$ with wall-normal position. The first row: the premultiplied streamwise spectra $k_x^+\phi_{uu}^+, k_x^+\phi_{vv}^+, k_x^+\phi_{ww}^+$ as functions of $z^+$ and $k_x^+$; the second row: the premultiplied spanwise spectra $k_y^+\Phi_{uu}^+, k_y^+\Phi_{vv}^+, k_y^+\Phi_{ww}^+$ as functions of $z^+$ and $k_y^+$. Contour lines show 1/6 to 5/6 of the maximum premultiplied energy, with an increment of 1/6. The shaded and solid contour lines show the results of original DNS data while the dash contour lines show the results of reconstructed velocity fields based on FLSE.

Figure 13 provides the full pre-multiplied spectra for original velocity fields and the FLSE-reconstructed ones, variating with the wall-normal position. The streamwise spectra $k_x\phi_{uu}, k_x\phi_{vv}, k_x\phi_{ww}$ are shown as functions of $z$ and $\lambda_x$ in the first row, and the spanwise spectra $k_y\Phi_{uu}, k_y\Phi_{vv}, k_y\Phi_{ww}$ are displayed as functions of $z$ and $\lambda_y$ in the second row. The energy spectra shown here are very consistent with the results of several investigators such as Hutchins & Marusic (2007) for $k_x\phi_{uu}$, Wang *et al.* (2017b) and Wang *et al.* (2018) for $k_y\Phi_{uu}$. The spectra from FLSE agree well with the original spectra for the bulk of the total energy. The performance for buffer layer ($z^+ < 80$) is excellent, recovering a satisfying energy distribution for almost the whole scale range of $\lambda_y < 2.7\delta$, $\lambda_x < 8\delta$ considered. The success in the buffer layer is attributed to the relatively smaller inducing range of vortices in this region, which is completely covered in the calculation domain for $\varphi_{ij}$ as shown in figure 8. For $z^+ > 80$, deviations between reconstructed spectra and the original spectra are observed for the large scales. Generally, the reconstructed spectra for $\phi_{vv}, \phi_{ww}$ is more accurate than that of $\phi_{uu}$, which is consistent with the fact that $u$ corresponds to comparatively larger coherence scales. A larger calculation domain of $\varphi_{ij}$ will account for the structures with larger scales embedded in the logarithmic but also significantly burden the calculation resourses, which is not involved in the work. All in all, the result of FLSE is satisfying for the scales range covered in the current calculation domain.

Energy spectra are among the primary considerations in modelling works of wall-bounded turbulence. To cover the whole scale range of energy spectra, Perry & Marusic (1995) proposed that three types of vortices need to be involved in the vortex model: two types of detached eddies accounting for the largest energy-containing motions and small-scale inertial motions respectively, and one type of the attached eddy accounting for the self-similar motions in middle scales. As for the representation of the so-called attached or detached eddies, kinds of controversy shows up. Earlier investigators treated the attached eddies or detached eddies as hierarchies of vortex tubes with $\Lambda$ or $\Omega$ shapes (Perry & Chong 1982) or vortex clusters (del Alamo *et al.* 2006) while recent works tend to recognize them by a pattern of velocities, such as the connected sets of points by Lozano-Durán & Jiménez (2014) and the iso-surface of $u$ by Lee *et al.* (2014) and Yoon *et al.* (2020). Marusic & Monty (2019) warned that the attached eddies do not account for the local enstrophy, and thus the detecting criterion should be independent of local $\nabla \mathbf{u}$. The current work does not contribute to the question of how to extract attached eddies from velocity fields or vortex fields. Instead, we attempt to estimate the upper limit of the accuracy for the vortex-based modelling works in predicting energy spectra. It shows that if the vortex fields are accurately modelled, it is possible to predict the bulk range of energy spectra, which is very promising for the vortex-based modelling works. Thus, we can see that although these fine-scale vortices identified based on the velocity gradient may not be suitable candidates for the attached eddies, they have great potential in turbulence modelling.

Another hot topic related to the energy spectra is the inner-outer interaction in wall-bounded turbulence. Plenty of investigations (Hutchins & Marusic 2007; Marusic *et al.* 2010; Mathis, Hutchins & Marusic 2011) have shown that large scale motions in the logarithmic layer have modulation and superposition effects on the velocity fluctuation in the near-wall region. And only the superposition effect contributes to the energy of the near-wall region, accounting for the increasing trend of inner peaks with the increase of Reynolds numbers. The superposition effect could be understood as the inducing effect of the vortices at the logarithmic layer, and FLSE allows us to have an insight into this aspect. In order to separate the inducing effects of vortices from different wall-normal positions on a reference plane, a V2V reconstruction based on height-filtered vortices is performed. Specifically, only the vortices located at a given height range $z \in [z_c - \Delta(z_c), z_c + \Delta(z_c)]$ are considered in the reconstruction of $u$ at the reference plane (e.g. $z^+ = 15.4$), and the result is denoted as $u(x, y; z_c)$. Herein, $z_c$ is the central position for the wall-normal range of the specified vortices; $\Delta(z_c)$ denotes the half-span, which is set as twice the local interval of the original wall-normal grid. Such vortex-filtered reconstruction is performed for a series of $z_c$ and the corresponding pre-multiplied streamwise spectra for $u(x, y; z_c)$ is calculated, which is denoted as $\phi_{uu}(k_x; z_c)$. Contours of $k_x z_c \phi_{uu}(k_x; z_c)/(2\Delta(z_c))$ are shown as functions of $k_x$ and $z_c$ in logarithmic coordinates in figure 14 (a), where the multiplier $z_c$ is added to balance the squeezing effect of logarithmic coordinates. This contour map shows the contributions of vortices at different wall-normal positions to $\phi_{uu}^+$ from $z^+ = 15.4$, which are called the contribution spectra. Similarly, the contribution spectra for $\phi_{vv}^+$, $\phi_{ww}^+$ from $z^+ = 15.4$ and the contribution spectra for $\phi_{uu}^+$, $\phi_{vv}^+$, $\phi_{ww}^+$ from $z^+ = 119.4$ are also calculated and displayed in the first and second row of figure 14, respectively.

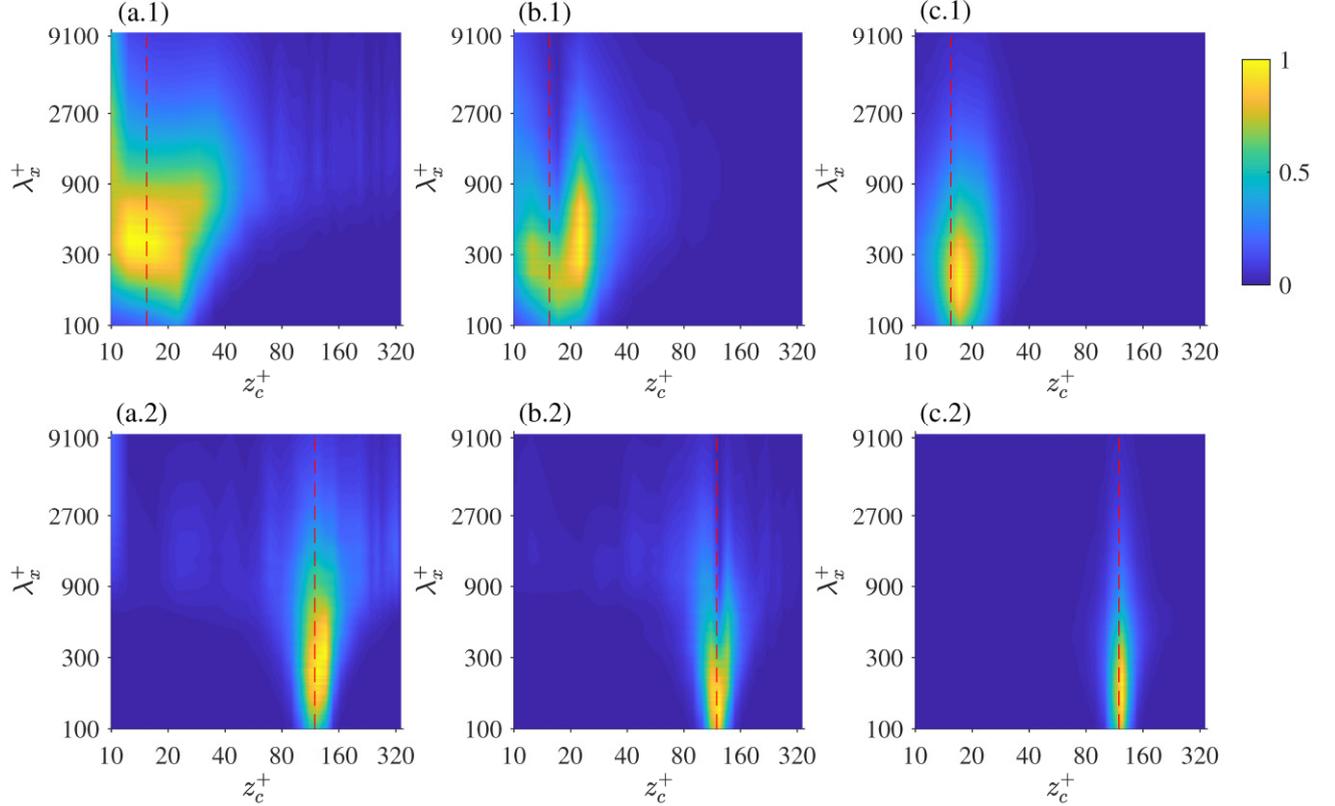

FIGURE 14. Contour maps showing the contributions of vortices at $z_c^+$ to $\phi_{uu}^+$, $\phi_{vv}^+$, $\phi_{ww}^+$ (a-c, respectively) from $z^+ = 15.6$ (the first row) and $z^+ = 119.6$ (the second row), as marked by the red dash lines. The contours show the local energy contribution premultiplied by $k_x^+ z_c^+$ and normalized by the maximum values.

Figure 14 is very revealing in several aspects. Firstly, it shows that compared to $\phi_{vv}(k_x; z_c)$ and $\phi_{ww}(k_x; z_c)$, the contribution spectra for $\phi_{uu}(k_x; z_c)$ is flatter along the $z_c$ direction, which indicates the $u$ component is more sensitive to the vortices below or above the reference positions. The sensitivity of $u$ to vortices may be owing to the mean velocity profile, which adds a strong wall-normal gradient of $u$. The instability triggered by vortices tends to produce low-speed structures such as the transient-growth mode in near-wall sustaining mechanics (Schoppa & Hussain 2002). In figure 14(a.1), the contribution from $z_c^+ > 100$ can be observed at $\lambda_x > 900$ for $\phi_{uu}(k_x; z_c)$, supporting the well-documented superposition effect of large scale structures on near-wall fluctuations. The energy contribution at the logarithmic layer does not show an obvious decreasing trend with increasing $z_c$. Thus, it can be inferred that the contributions from the outer region would increase with the thickness of the boundary layer ($\delta^+$), or the Reynolds number equivalently. Different from $\phi_{uu}(k_x; z_c)$, the main energy-contributing regions for $\phi_{vv}(k_x; z_c)$ and $\phi_{ww}(k_x; z_c)$ are limited below $z_c^+ < 100$ and $z_c^+ < 50$ respectively. For the contribution spectra of $z^+ = 119.4$, while the energy-contributing regions for $\phi_{uu}(k_x; z_c)$ and $\phi_{vv}(k_x; z_c)$ covers the whole wall-normal range considered, the energy-contributing region for $\phi_{ww}(k_x; z_c)$ remains in the local. The special behaviour of $w$ component has also been noticed in the scaling law of $\langle w^2 \rangle$, which keeps a constant in the logarithmic region while both $\langle u^2 \rangle$ and $\langle v^2 \rangle$ shows a logarithmic law with the increase of wall-normal position (Woodcock & Marusic 2015). The special behaviour of the $w$ component is owing to the bounding effect of the wall, which significantly weakens the inducing effect of vortices on the $w$ component. This explanation stems from the hypothesis of the attached eddy theory, which claims that the modelled energy-containing eddies should feel the existence of the wall. In fact, the requirement that induced flow by attached eddies cannot penetrate through the wall is imposed as a boundary condition to deduce the scaling law for $\langle u^2 \rangle, \langle v^2 \rangle, \langle w^2 \rangle$ in Perry & Marusic (1995) and Woodcock & Marusic (2015). Figure 14(c.1) provide evidence for this condition, and this revealing result is attributed to the FLSE developed in this work.

*5.5 High-order moments of FLSE-reconstructed TBL*

The energy spectra discussed above only involve the second-order moments of velocities. The higher-order moments offer further indicators to distinguish the reconstructed velocity fields and original DNS velocity fields. This subsection focuses on higher-order moments, including the skewness and flatness for the three velocity components. The skewness and flatness for $u$ (denoted as $S_u$ and $F_u$) are defined as $\langle u^3 \rangle / \langle u^2 \rangle^{3/2}$ and $\langle u^4 \rangle / \langle u^2 \rangle^2$, respectively. And these results are shown in figure 15 with the second-order moments also provided as references. It shows that the second-order moments from FLSE are lower than the original DNS results, which is consistent with the observations on the energy spectra in figure 12 and figure 13. As for the skewness and flatness of $u$, the results from FLSE shows distinctive behaviours at the near-wall region compared to the original DNS results. For the FLSE results from $15 < z^+ < 100$, the skewness is positive, and the flatness is larger than three, both of which are on the opposite of the corresponding DNS results. A positive skewness indicates that the extreme events of $u$ corresponding to a large fluctuation magnitude are more likely positive, and vice versa. And a flatness of three is a behaviour of standard Gaussian distribution and larger or smaller flatness corresponds to a super-Gaussian or sub-Gaussian behaviour, respectively. It can be inferred from the deviations that the reconstruction $u$ tends to be flatter and contain weak negative fluctuations compared to the original field. In the logarithmic region, the deviations for both skewness and flatness are reduced: both results show a negative skewness and sub-Gaussian behaviour. Different from the results of $u$, the skewness and flatness of $v, w$ from FLSE agree well with the original DNS data for the whole wall-normal range considered, which further validates that $u$ pose the most significant challenge for accurate reconstruction.

Skewness and flatness have also been considered in the attached eddy model. The model predicts that the fourth-order moment, i.e. flatness of $u$, is invariably larger than three (Woodcock & Marusic 2015), which is different from the well-accepted experimental results of 2.8 (Fernholz & Finleyt 1996; Stanislas *et al.* 2008). To build on this aspect, de Silva *et al.* (2016b) introduced the spatial exclusion effect in attached eddy model, which was implemented by imposing a limit on the minimum distance of attached eddies. They found that the improved attached eddy model gives a correct prediction on the flatness of $u$, which validates the importance of the spatial exclusion effect in deriving accurate predictions. The current work reconstructs the velocity fields based on the ground truth vortex fields, which embedded the spatial exclusion effects naturally. The reconstruction results retain the sub-Gaussian behaviour of $u$ for the logarithmic layer but not for the near-wall region. Thus, it can be inferred that some effects alternative to the spatial exclusion and neglected by FLSE distorted the flatness in the near-wall region.

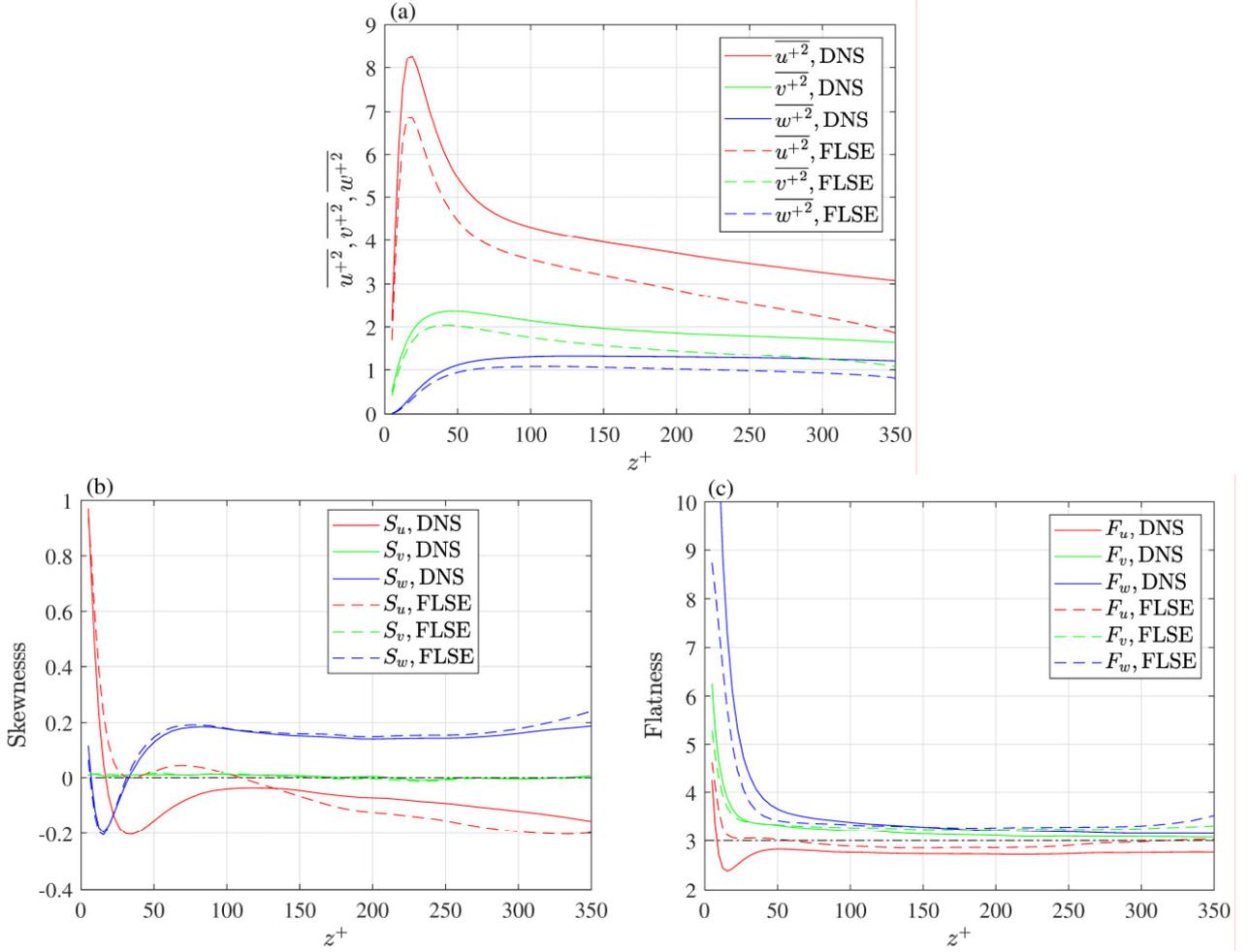

FIGURE 15. The second to fourth moments for $u^+, v^+, w^+$ as functions of wall-normal positions. (a) Fluctuation energy, (b) skewness, (c) flatness.

In order to reveal the reason that FLSE fails to recover the skewness and flatness in the near-wall region, the inducing model employed by FLSE needs to be reconsidered. FLSE expressed the inducing effect of vortices as linear operators, which implies that the inducing effects of two vortices with the opposite orientations can be modelled by one function with adjustable signs or multipliers. However, the real situation might be more complex, sometimes beyond the capacity of the linear operators. To prove this argument, figure 16 shows conditionally-averaged $u$ field around a given positive streamwise vortex at $z^+ = 37.8$, which are calculated by $\langle u(\mathbf{r} + \Delta\mathbf{r})\Lambda_x^p(\mathbf{r})\rangle_{\mathbf{r},t}$. Herein, the superscript $p$ indicates a positive-pass filter which retains positive values of the input field and sets the negative values as zeros. The wall-normal position for the conditional average corresponds to the largest deviation of the skewness for reconstructed $u$ in figure 15 (b). As we can see, the $u$ field around a positive vortex shows an asymmetric pattern. In this contour map, the upper side of the central vortex corresponds to negative $u$ with larger magnitude, which is consistent with the fact that a positive streamwise vortex distributes below a low-speed streak. In the implementation of FLSE, the inducing effects of a positive streamwise vortex and a negative streamwise vortex are averaged in one correlation function (as shown in figure 8), which results in a symmetric pattern for inducing models. Such averaging operation will reduce the intensity of low-speed streaks in reconstructed fields and result in a smaller skewness magnitude and a larger flatness, which satisfyingly explains the deviations of the FLSE results displayed by figure 15. To deal with this issue, the reconstruction scheme should be improved by distinguishing the inducing effects of vortices with different signs, which is expected to be the topic of future work.

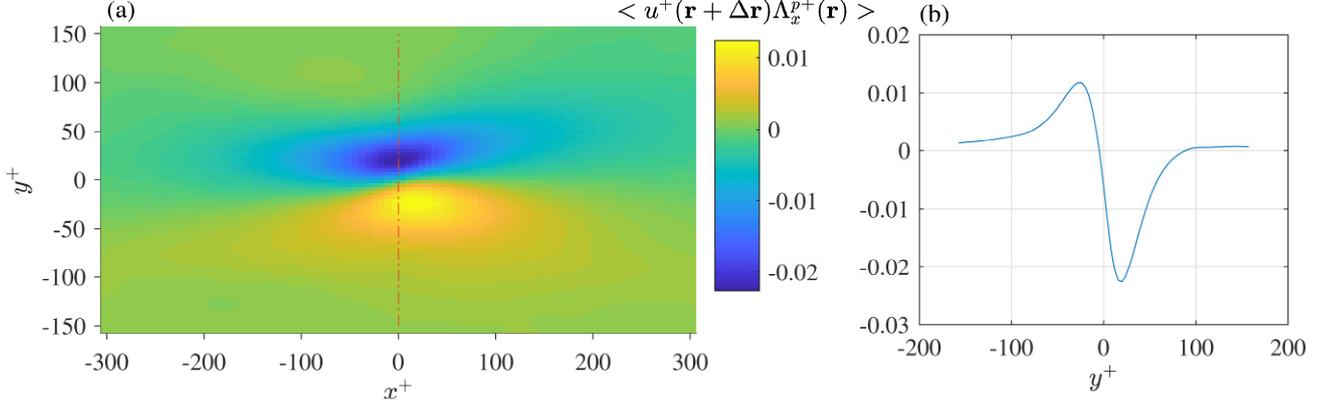

FIGURE 16. Conditionally-averaged $u^+$ field around a given positive streamwise vortex at $z^+ = 37.8$. (a) Contour map of the conditionally-averaged $u^+$ field, (b) the conditionally-averaged $u^+$ as a function of $y^+$ along a line extracted from the contour map, as marked by the red dash-dot line.

## 6  Concluding remarks

Representing the turbulent motions by evolving vortex structures is a fundamental viewpoint in wall-bounded turbulence. The current work attempted to answer three questions correlated with the vortex-representing view: what is the relationship between vortex magnitude and vortex orientation, how to transform a vortex field into the corresponding velocity field and how much turbulence is determined by vortices. The investigations were based on the DNS data, which allowed sizeable statistical range and suffered no resolution or uncertainty issues.

The relationship between vortex orientation and magnitude is quantitatively investigated by the differential geometry, which is an essential mathematical tool to represent the local curvatures of a surface. The vortex vector was projected onto the local curvature coordinate system of the vortex surface, and the resulting spherical PDF showed a clear trend of aligning with the first principal direction of the local curvatures. It validated that the vortex orientation tended to align with the slowest-changing direction of the vortex magnitude. Furthermore, the divergence of the vortex field was investigated, and it showed that the vortex field could be regarded as a divergence-free field with small deviations. The divergence-free property and the relationship between vortex magnitude and orientation were combined in an attempt to infer vortex components based on the vortex magnitude, and the results were promising. The properties revealed by this work validated that the vortex structures identified from $\lambda_{ci}$ and $\Lambda_r$ are good candidates for modelling objects in wall-bounded turbulence, which adds value to the following reconstruction works.

The problem of vortex-to-velocity reconstruction was discussed in the general theoretical framework of linear stochastic estimation. The reconstruction was carried out by using linear operators, which needed to be solved from a set of governing equations. The governing equation incorporated the correlation information of vortex fields and velocity fields, which could be significantly simplified by using the homogenous condition of wall-bounded turbulence. The newly-developed method was named as the field-based linear stochastic estimation. The FFT implementation and correlated considerations were introduced, which could be viewed as an expansion of SLSE form 1D to 3D application in wall-bounded turbulence. The roles of the regularization parameter and the Hanning window involved in the FFT implementation were revealed based on a test on the DNS data. It showed that Hanning window was necessary for numerical instability and smaller regularization parameter corresponded to higher accuracy but worse statistical convergence for the solution.

The performance of FLSE in reconstructing velocity fields was compared with the Biot-Savart law from several aspects. For the instantaneous reconstructed fields, FLSE satisfyingly recovered the flow structures in both the region close to the wall and the logarithmic region while BS law failed to recover the low-speed streaks very close to the wall. The excellent performances of FLSE were also observed in the correlation coefficients, energy spectra and also the linear coherent spectra, which validated the effectiveness of FLSE. Mostly, FLSE performed better than BS law except for the large scales in the logarithmic region truncated by the calculation domain. The success of FLSE was attributed to the empirical model functions, which were driven from the DNS data in a minimum deviation sense.

The streamwise and spanwise spectra for all the velocity components from the reconstructed fields were compared with the original DNS spectra. Generally speaking, the spectra from FLSE were reasonably consistent with the original DNS spectra for the bulk of total energy. The most significant deviations occurred in the outer region for $u$ component, which was caused by the limited calculation domain for the inducing model functions. The results indicated that if the vortex fields had been accurately modelled, the energy spectra could be recovered for the bulk of the scale range, which was very promising the vortex-based modelling works. To shed light on the inner-outer interaction, the energy spectra were further decomposed according to the contributions from different wall-normal positions. It showed that the spectra of $\phi_{uu}$ was most sensitive to the vortices below or above the reference plane, which was correlated with the embedded velocity gradient in TBL. The results supported the previous viewpoints that the large scale motions at the logarithmic region have a superposition effect on the fluctuations in the inner layer. Furthermore, it showed that the energy contribution region for $\phi_{ww}$ was limited at the neighbouring region of the reference height, which was linked to the special behaviour of $\langle w^2 \rangle$ in the scaling law.

At last, high-order moments for $u, v, w$ from reconstructed fields were compared with the original DNS data. While the skewness and flatness for $v, w$ from FLSE agreed well with the original DNS data, large deviations occurred for $u$ at the near-wall region. The reason was revealed by analyzing the conditionally-averaged velocity field around a positive streamwise vortex, which presented an asymmetric pattern. However, in the implementation of FLSE, the inducing effects of positive streamwise vortices and negative streamwise vortices were averaged in one correlation function, which resulted in a symmetric pattern for inducing models. It was expected that FLSE might be further improved by distinguishing the inducing effects of vortices with opposite orientations in future work.

**Acknowledgements**

This work was supported by the National Natural Science Foundation of China (grant 11490552, 11902371, 91852204) and the Fundamental Research Funds for the Central Universities (2019QNA4056).

**Appendix A: calculation for principal directions of local curvatures**

Following the conventions in this article, the vortex magnitude field is designated by $\lambda_{ci}(\mathbf{r})$. Focus on the neighbourhood of a given point $\mathbf{r}_0$, and let $\delta \mathbf{r} = \mathbf{r} - \mathbf{r}_0$ denote the relative position vector regarding $\mathbf{r}_0$ ($|\delta \mathbf{r}| = |\mathbf{r} - \mathbf{r}_0| \ll 1$). According to Taylor's theorem, $\lambda_{ci}(\mathbf{r})$ could be approximated as a quadrant form, as

$$\lambda_{ci}(\mathbf{r}) = \lambda_{ci}(\mathbf{r}_0) + \delta \mathbf{r}^T \mathbf{G} + \frac{1}{2} \delta \mathbf{r}^T \mathbf{H} \delta \mathbf{r} + \mathrm{O}(\delta r^2), \tag{A 1}$$

where $\mathbf{G} = \nabla \lambda_{ci} = \left[ \dfrac{\partial \lambda_{ci}(\mathbf{r}_0)}{\partial x}, \dfrac{\partial \lambda_{ci}(\mathbf{r}_0)}{\partial y}, \dfrac{\partial \lambda_{ci}(\mathbf{r}_0)}{\partial z} \right]^T$ denotes the gradient of $\lambda_{ci}(\mathbf{r})$, and

$$\mathbf{H} = \begin{bmatrix} \dfrac{\partial^2 \lambda_{ci}(\mathbf{r}_0)}{\partial x^2} & \dfrac{\partial^2 \lambda_{ci}(\mathbf{r}_0)}{\partial x \partial y} & \dfrac{\partial^2 \lambda_{ci}(\mathbf{r}_0)}{\partial x \partial z} \\ \dfrac{\partial^2 \lambda_{ci}(\mathbf{r}_0)}{\partial x \partial y} & \dfrac{\partial^2 \lambda_{ci}(\mathbf{r}_0)}{\partial y^2} & \dfrac{\partial^2 \lambda_{ci}(\mathbf{r}_0)}{\partial y \partial z} \\ \dfrac{\partial^2 \lambda_{ci}(\mathbf{r}_0)}{\partial x \partial z} & \dfrac{\partial^2 \lambda_{ci}(\mathbf{r}_0)}{\partial y \partial z} & \dfrac{\partial^2 \lambda_{ci}(\mathbf{r}_0)}{\partial z^2} \end{bmatrix}, \tag{A 2}$$

which is the so-called Hessian matrix.

The principal directions for local curvatures should be searched in the tangent plane of local vortex surface. Thus we impose a constrain that $\delta \mathbf{r}$ should be a vector in the tangent plane of the local vortex surface, which means

$$\mathbf{G}^T \delta \mathbf{r} = 0. \tag{A 3}$$

Any $\delta \mathbf{r}$ satisfying (A 3) could be expressed as

$$\delta \mathbf{r} = \alpha_1 \mathbf{e}_1 + \alpha_2 \mathbf{e}_2 = [\mathbf{e}_1, \mathbf{e}_2][\alpha_1, \alpha_2]^T = \mathbf{E} \boldsymbol{\alpha}, \tag{A 4}$$

where $\mathbf{e}_1, \mathbf{e}_2$ are two unit orthogonal vectors satisfying

$$\mathbf{G}^T \mathbf{e}_1 = \mathbf{G}^T \mathbf{e}_2 = 0, \tag{A 5}$$

and $\boldsymbol{\alpha}$ could be any real vector indicating a direction in the tangent plane.

Under this constrain of (A 3) and employing the representation of (A 4), it can be derived that

$$\lambda_{ci}(\mathbf{r}) = \lambda_{ci}(\mathbf{r}_0) + \frac{1}{2} \boldsymbol{\alpha}^T \mathbf{E}^T \mathbf{H} \mathbf{E} \boldsymbol{\alpha} + \mathrm{O}(\delta r^2). \tag{A 6}$$

Now consider the variation of $\lambda_{ci}$ along one direction in the tangent plane of the vortex surface, which can be achieved from (A 6) by linearly changing $\boldsymbol{\alpha}$, i.e. $\boldsymbol{\alpha} = \boldsymbol{\alpha}_0 t$. Herein, $\boldsymbol{\alpha}_0$ is a given unit direction vector, and $t$ is a variable parameter. Obviously, $\lambda_{ci}$ varies with $t$ as a parabolic curve. The principal directions correspond to the directions with the largest and smallest curvatures for $\lambda_{ci}(t)$, which is equivalent to the eigenvalue problem of $\mathbf{E}^T \mathbf{H} \mathbf{E}$. Let $\mathbf{v}_1$ and $\mathbf{v}_2$ denote the eigenvectors for $\mathbf{E}^T \mathbf{H} \mathbf{E}$, corresponding to two eigenvalues of $\mu_1, \mu_2 (\mu_1 > \mu_2)$, respectively. Then, the first and second principal directions for the local curvatures at $\mathbf{r}_0$ should be $\delta \mathbf{r}_1 = \mathbf{E} \mathbf{v}_1$ and $\delta \mathbf{r}_2 = \mathbf{E} \mathbf{v}_2$.

**Appendix B: deduce equation (4.4) from equation (4.3)**

For simplicity, rewrite the deviation $D$ in the form of inner product, i.e.

$$D = \iiint_\Omega (\hat{u}_i - u_i)(\hat{u}_i - u_i) d\Omega / \iiint_\Omega d\Omega = (\hat{u}_i - u_i, \hat{u}_i - u_i) \tag{B 1}$$

where $(\cdot, \cdot)$ represents the inner product of the two terms separated by the comma. The repeated index $i$ of the two bracketed terms implies an automatic summation, following Einstein's summation convention.

To find $L_{ij}$ which results in the minimum $\langle D \rangle$, give a small variation ($\delta$) on $L_{ij}$ and investigate the corresponding variation of $\langle D \rangle$. Thus we have

$$\delta \langle D \rangle = \delta \langle (\hat{u}_i - u_i, \hat{u}_i - u_i) \rangle = 2 \langle (\hat{u}_i - u_i, \delta \hat{u}_i) \rangle. \tag{B 2}$$

Note that the implicit position vector is $\mathbf{r}$ for $\hat{u}_i, u_i, \delta \hat{u}_i$. According to the calculus of variations, the minimum of $\langle D \rangle$ corresponds to a balanced position, where for any possible variation of $L_{ij}$, the corresponding variation of $\langle D \rangle$ should be zero, i.e.

$$\delta \langle D \rangle = 2 \langle (\hat{u}_i - u_i, \delta \hat{u}_i) \rangle = 0 \quad \text{for any possible} \quad \delta \hat{u}_i = \delta(L_{ij}) \Lambda_j. \tag{B 3}$$

Now, let $\delta \hat{u}_i = \delta L_{ij} \Lambda_j = \Lambda_m(\mathbf{r}') \delta(\mathbf{r} - \mathbf{r}'') \delta_{in}$, where $\delta(\mathbf{r} - \mathbf{r}'')$ is the 3D Dirac delta function and $\delta_{in}$ is the three-order Kronecker delta. $\mathbf{r}', \mathbf{r}''$ could be any reference position vectors and $m, n$ are variable indices without limitation. The specific definitions for $\delta(\mathbf{r} - \mathbf{r}'')$ and $\delta_{in}$ could be indicated by

$$(f(\mathbf{r}), \delta(\mathbf{r} - \mathbf{r}'')) = \int_\Omega f(\mathbf{r}) \delta(\mathbf{r} - \mathbf{r}'') d\Omega = f(\mathbf{r}''), \tag{B 4}$$

where $f(\mathbf{r})$ could be any 3D continuous function, and

$$\delta_{in} = \begin{cases} 0 & \text{if } i \neq n \\ 1 & \text{if } i = n \end{cases}. \tag{B 5}$$

The prescription for the form of $\delta \hat{u}_i$ is based on two considerations. First, the form should facilitate an effective simplification for the form of $\delta \langle D \rangle$. Second, the form should have plenty of variations by adjusting the embedded indices or parameters, in order to obtain enough equations for the unknown $L_{ij}$.

$\delta \hat{u}_i = \delta L_{ij} \Lambda_j = \Lambda_m(\mathbf{r}') \delta(\mathbf{r} - \mathbf{r}'') \delta_{in}$ implies a possible variation for $L_{ij}$, which results in a variation of $\langle D \rangle$ as

$$\delta \langle D \rangle = 2 \langle (\hat{u}_i - u_i, \Lambda_m(\mathbf{r}') \delta(\mathbf{r} - \mathbf{r}'') \delta_{in}) \rangle = 0. \tag{B 6}$$

Simplifying this equation by using (B 4) and (B 5) gives

$$\langle (\hat{u}_i - u_i, \Lambda_m(\mathbf{r}') \delta(\mathbf{r} - \mathbf{r}'') \delta_{in}) \rangle = \langle (\hat{u}_n(\mathbf{r}'') - u_n(\mathbf{r}'')) \Lambda_m(\mathbf{r}') \rangle = 0. \tag{B 7}$$

In the following equations, $\mathbf{r}''$ is neglected for simplicity. Considering that $\hat{u}_n = L_{nj} \Lambda_j$, we have

$$\langle (L_{nj} \Lambda_j - u_n) \Lambda_m(\mathbf{r}') \rangle = \langle L_{nj} \Lambda_j \Lambda_m(\mathbf{r}') \rangle - \langle u_n \Lambda_m(\mathbf{r}') \rangle = 0. \tag{B 8}$$

According to the linear operation of $L_{nj}$, finally we get

$$L_{nj}\langle \Lambda_j \Lambda_m(\mathbf{r}')\rangle = \langle u_n \Lambda_m(\mathbf{r}')\rangle. \tag{B 9}$$

(B 9) is equivalent to (4.4) in the article by replacing the free index $n$ with $i$. What is noteworthy is the equation holds for all the possible $\Lambda_m(\mathbf{r}'), m=1,2,3; \mathbf{r}' \in \Omega$, number of which is large enough to determine $L_{nj}$.